\documentclass[twocolumn, notitlepage, 10pt, aps, floatfix, showpacs, prb, citeautoscript, superscriptaddress, longbibliography]{revtex4-1}
\usepackage{graphicx}
\usepackage{amsmath, amssymb}
\usepackage{bm}
\usepackage{xcolor}
\usepackage[colorlinks, citecolor=blue, urlcolor=blue, linkcolor=red]{hyperref}
\usepackage{bookmark}
\usepackage[version=4]{mhchem}
\usepackage{microtype}
\usepackage{float}
\usepackage[load=physical,load=abbr, range-units=single]{siunitx}

\begin{document}
\title{Quality factor for zero-bias conductance peaks in Majorana nanowire}
	
\author{Yi-Hua Lai}
\author{Sankar Das Sarma}
\author{Jay D. Sau}
\affiliation{Department of Physics, Condensed Matter Theory Center and the Joint Quantum Institute, University of Maryland, College Park, Maryland 20742, USA}

\begin{abstract}
Despite recent experimental progress towards observing large zero-bias conductance peaks (ZBCPs) as signatures of Majorana modes, confusion remains about whether Majorana modes have been observed. This is in part due to the theoretical prediction of fine-tuned trivial (i.e., non-topological) zero-bias peaks that occur because of uncontrolled quantum dots or disorder potentials. While many aspects of the topological phase can be somewhat fine-tuned because the topological phase space is often small, the quantized height of the ZBCP associated with a Majorana mode is known to be robust at sufficiently low temperatures even as the tunnel barrier is pinched off to vanishingly small normal-state conductance. The key shortcoming of the existing experimental works is an acute lack of stability of the putative Majorana mode features, indicating the probable absence of a topological phase, and the current paper suggests specific experimentally accessible measures for a careful quantitative analysis of the measured ZBCP stability. In this paper, we study how the counter-intuitive robustness of the ZBCP height to the tunnel barrier strength can be used to distinguish Majorana modes from non-topological ZBCPs. To this end, we introduce a dimensionless quality factor $F$ to quantify the robustness of the ZBCP height based on the range of normal-state (i.e. above-gap) conductance (which depends crucially on the tunnel barrier height) over which the ZBCP height remains within a pre-specified range of quantization. By computing this quality factor $F$ together with the topological characteristics for a wide range of models and parameters, we find that Majoranas are significantly more robust (i.e., have a higher value of $F$) compared with non-topological ZBCPs in the ideal low-temperature limit. Even at a temperature as high as the experimentally used $20$ mK, we find that we can set a threshold value of $F\sim 2.5$ (for $\epsilon=0.1$) so that ZBCPs associated with a quality factor $F>2.5$ are likely topological and $F\ll 2.5$ are topologically trivial. More precisely, the value of $F$ is operationally related to the degree of separation of the Majorana modes in the system, although $F$ uses only the experimentally measured tunnel conductance properties. Finally, we discuss how the quality factor $F$ measured in a transport setup can help estimate the quality of topological qubits made from Majorana modes. In particular, we show that if the induced gap can be enhanced somehow beyond the currently available $\sim 30$ $\micro$eV in InAs/Al samples, large (small) values of $F$ could easily distinguish between stable topological (unstable trivial) ZBCPs with the quantum dot induced quasi-Majorana bound states occasionally behaving similar to topological Majorana modes in short wires.
\end{abstract}
	
\maketitle
	
\section{Introduction}\label{sec:level1_1}
Since the model of a Majorana nanowire [one-dimensional superconductor-semiconductor (SC-SM) heterostructure with the $s$-wave superconducting proximity effect, spin-orbit coupling, and spin-splitting by Zeeman field] was proposed in 2010~\cite{Lutchyn2010Majorana,Sau2010Generic,Sau2010NonAbelian,Oreg2010Helical}, Majorana zero modes (MZMs), which are the non-Abelian topological quantum computing qubits~\cite{Kitaev2001Unpaired,Kitaev2003Faulttolerant,Freedman2003Topological,DasSarma2005Topologically,Nayak2008NonAbelian}, have been broadly studied in the experiments over the past decade~\cite{Mourik2012Signatures,Das2012Zerobias,Deng2012Anomalous,Churchill2013Superconductornanowire,Finck2013Anomalous,Deng2016Majorana,Nichele2017Scaling,Zhang2017Ballistic,Chen2017Experimental,Gul2018Ballistic,Kammhuber2017Conductance,Vaitiekenas2018Effective,Moor2018Electric,Bommer2019SpinOrbit,Grivnin2019Concomitant,Chen2019Ubiquitous,Anselmetti2019Endtoend,Menard2020ConductanceMatrix,Puglia2021Closing,Yu2021NonMajorana,Zhang2021Large}. Especially, the observations of zero-bias conductance peaks (ZBCPs), one of the most important signatures of MZMs, have been widely reported in the tunneling spectroscopy of InAs or InSb nanowires~\cite{Mourik2012Signatures,Das2012Zerobias,Deng2012Anomalous,Churchill2013Superconductornanowire,Finck2013Anomalous,Deng2016Majorana,Nichele2017Scaling,Zhang2017Ballistic,Vaitiekenas2018Effective,Moor2018Electric,Zhang2018Quantized,Bommer2019SpinOrbit,Grivnin2019Concomitant,Anselmetti2019Endtoend,Menard2020ConductanceMatrix,Puglia2021Closing,Zhang2021Large}. In fact, ZBCPs with heights near the theoretically predicted quantized value of $2e^2/h$ have been seen in experiments~\cite{Nichele2017Scaling,Zhang2018Quantized,Zhang2021Large}, leading to optimism regarding the observation of Majoranas.
	
Unfortunately, several mechanisms for ZBCPs whose conductance height may be tuned to be near the quantized value have also been identified theoretically since the original observation of ZBCPs in Majorana nanowire systems. For instance, Andreev bound states (ABSs) induced by quantum dots (QDs) or inhomogeneous chemical potential can deterministically produce quantized ZBCPs~\cite{Pan2020Physical, Kells2012Nearzeroenergy,Liu2012ZeroBias,Prada2012Transport,Liu2017Andreev,Moore2018Quantized,Moore2018Twoterminal,Vuik2019Reproducinga} (``bad" ZBCPs in the terminology introduced in Ref.~\cite{Pan2020Physical}). Alternatively, disorder-induced random potential can with sufficient fine-tuning create the trivial ZBCPs quantized near $2e^2/h$~\cite{Bagrets2012Class,Pikulin2012zerovoltage,Sau2013Density,Mi2014Xshaped,Pan2020Physical} (``ugly" ZBCPs as called in Ref.~\cite{Pan2020Physical}) as well. Sometimes, these topologically-trivial subgap bound states can even accidentally display the stable quantized conductance~\cite{Pan2020Generic,DasSarma2021Disorderinduced,Pan2021Crossover,Pan2021Quantized,Pan2021Ondemand} to some extent when the system is fine-tuned, leading to the possibility of misconstruing such trivial zero-energy bound states in experiments as MZMs, as has been done repeatedly in the literature. This leads to the central challenge in the field, i.e., distinguishing topological MZMs from such trivial zero-energy fermionic bound states based on the currently feasible experimental techniques. The claimed `quantization' in this experiment~\cite{Zhang2018Quantized} is now understood to be an unfortunate confirmation-biased outcome of fine-tuning and post-selection~\cite{Zhang2021Large}. This has led to a justified retraction of the original Ref.~\cite{Zhang2018Quantized}, but we believe that all existing local tunnel spectroscopic experimental ZBCP-based claims of Majorana observations in the literature most likely are untenable because the robustness of the ZBCP (the subject matter of the current paper) and their generic nonlocal nature has never been established. Clearly, just observing ZBCPs with tunnel conductance $\sim 2e^2/h$ is insufficient evidence for MZMs. The importance of the stability of quantized ZBCPs motivates us to come up with a quantity that can measure the robustness of quantization.
	
It is indeed true that one of the most striking necessary properties that has been predicted for MZMs is the quantization of the ZBCP at zero temperature. Unlike the mere presence or absence of a zero-energy state, which turns out to be determined even in the non-topological case by fine-tuning~\cite{Kells2012Nearzeroenergy,Liu2012ZeroBias,Bagrets2012Class,Pikulin2012zerovoltage,Prada2012Transport,Sau2013Density,Mi2014Xshaped,Liu2017Andreev,Moore2018Quantized,Moore2018Twoterminal,Vuik2019Reproducinga,Kells2012Nearzeroenergy,Liu2017Andreev,Liu2018Distinguishing,Moore2018Twoterminal,Vuik2019Reproducinga,Pan2020Physical}, the quantization of the peak height is a quantitative signature that can allow us to associate a Majorana with a qualitative property as opposed to a mere presence or absence of a tunneling peak. We introduce a quality factor to characterize this property, which should be measurable in experiments. This quality factor that can be assessed from a transport measurement, ideally, is connected with the decoherence time of a topological qubit that would be constructed from this Majorana device~\cite{Sau2020Counting}. The quantization of the peak height is striking because it is robust, in principle, to the tunnel barrier height even when it is very large where the barrier has a negligible transmission probability for electrons. While this Majorana quantization independent of the barrier height has some similarity to resonant transmission through a symmetric double barrier potential, the quantization in the Majorana case is protected by particle-hole symmetry of the superconductor, which cannot be lifted by an external perturbation, making the quantization exact. This motivates the question of whether the quantization of the ZBCP height can be used to separate out topological Majorana states from other non-topological ZBCPs in some indirect manner going beyond just measuring the ZBCP magnitude. While recent experiments have seen large ZBCPs, very few have seen heights close to the quantized value. Even in the few reported quantized cases, the parameter range for the robustness is quite small. The robustness with respect to most parameters such as magnetic field or gate voltages, which are dimensionful, are hard to quantify as stable because the comparison standard is not obvious while varying a dimensionful parameter with a unit. Also variations in these experimental parameters may affect the topological phase of the system, particularly if the topological phase is fragile in parameter space as is often the case. In contrast, variations in the tunnel barrier can be quantified by the dimensionless normal-state conductance (i.e., relative to the conductance quantum), which cannot affect the bulk topological properties. Thus, the ZBCP height in the topological case should remain quantized as long as the normal-state conductance exceeds a limit proportional to the temperature. This motivates our introduction of the quality factor concept to characterize Majorana modes through tunneling spectroscopy. The basic idea is that not all experimentally tunable parameters are equivalent: While the applied magnetic field and gate voltages directly affect the topological phase diagram by controlling the spin splitting and the chemical potential, the tunnel barrier and the temperature are fundamentally different in controlling the quantitative aspects of the Majorana quantization without affecting the topological phase itself.
	
The above discussion of the quantization of the ZBCP height in the case of the ideal Majorana raises the central question asked in this paper, i.e., how does the fine-tuned apparent robustness of the ZBCP height in non-topological ZBCPs compare to the intrinsic protected robustness of the topological Majorana. Given that the tunnel barrier robustness of the Majorana ZBCP depends on temperature, we expect the comparison to depend on temperature as well since it is well-established that the Majorana ZBCP quantization depends on tunnel barrier and temperature in an intrinsically coupled manner~\cite{Sengupta2001Midgapa,Setiawan2017Electron}. While we will present results for a range of temperature, we will focus our discussion at $20$ mK as the lowest practical temperature based on measurement reports so far. It will become clear later in this paper how crucial it is to carry out measurements at the lowest possible temperatures by virtue of the fact that the realistic topological superconducting gap in currently available nanowires is very small. One of the key results we will discuss are plots of the ZBCP height versus the normal-state conductance through the tunnel barrier. While these plots are the regular way to present the measured conductance in the experiments~\cite{Zhang2018Quantized, Zhang2021Large}, no earlier theoretical work systematically calculates the ZBCP versus the normal-state conductance in detail. We use these plots to quantify a measure of the robustness, i.e., a dimensionless quality factor, which enables us to assign a single precisely defined number $F$ to the ZBCP at a particular temperature. We will find that the quality factor $F$ associated with robustness to the tunnel barrier tuning can distinguish between topological and non-topological ZBCPs using a threshold value for $F$. To draw this conclusion, we study the spatial separation of the Majorana components of the lowest wave function to determine if a particular set of model parameters is in the topological superconducting phase. While transport experiments do not have access to the Majorana wave-function profiles and bulk gap closing and reopening phenomenon near topological quantum phase transition (TQPT) is likely to be too weak for realistic long topological wires to experimentally identify~\cite{Pan2021Threeterminal}, we should use the newly-defined quality factor $F$, which is determined by the measured conductance, to assess whether a particular ZBCP in experiment is topologically non-trivial. An experimental determination of $F$ would, therefore, provide strong support for the existence or not of topological MZMs in a given sample.

We note that our current paper may in some sense be construed as a quantitative data analysis tool for future Majorana tunneling measurements observing ZBCPs. The key advantage of our theoretical proposal is that it requires only measured quantities, namely, the ZBCP itself and the above-gap normal conductance as a function of the tunnel gate voltage. Since we focus quantitatively on the currently used InAs/Al semiconductor-superconductor hybrid systems (with its rather small induced gap), our results are applicable directly to the ongoing local tunneling spectroscopy experiments in many laboratories without requiring more complex three-terminal setups for nonlocal conductance measurements. In fact, nonlocal measurements (and then braiding measurements) should only be carried out on those samples which satisfy the stability protocol worked out in the current paper, saving a great deal of time and effort in future experiments. We emphasize that the lack of stability of the ZBCP quantization has been the key issue preventing progress in the field, and our suggested precise and quantitative protocol based only on local tunneling measurements should help future measurements to determine a zero-th order distinction between topological and trivial ZBCPs without resorting to extensive (and always somewhat unreliable, since the sample parameters and the level of disorder are never known accurately) simulations as our quality factor is determined uniquely by measured quantities with no need for any theoretical simulations.
		
The rest of this paper is organized as follows. In Sec.~\ref{sec:level1_2}, we describe our theoretical model by introducing the Hamiltonion of the SC-SM nanowire, various potentials for the scenarios we study in this paper, Majorana-composed wave functions, and the formalism for the quality factors. In Sec.~\ref{sec:level1_3}, we show our numerical results for the cases of ``good", ``bad", and ``ugly" ZBCPs~\cite{Pan2020Physical}. We show conductance color plots and conductance line cuts as a function of magnetic field and tunneling barrier height, Majorana-composed wave functions, ZBCP as a function of normal-metal conductance, and quality factors as a function of temperature in each panel. In Sec.~\ref{sec:level1_4}, we discuss some generic features from our numerical results using a panoramic angle. Finally, we make a conclusion in Sec.~\ref{sec:level1_5} with a summary.
	
\section{Model}\label{sec:level1_2}
\begin{figure}
	\includegraphics[scale=0.5]{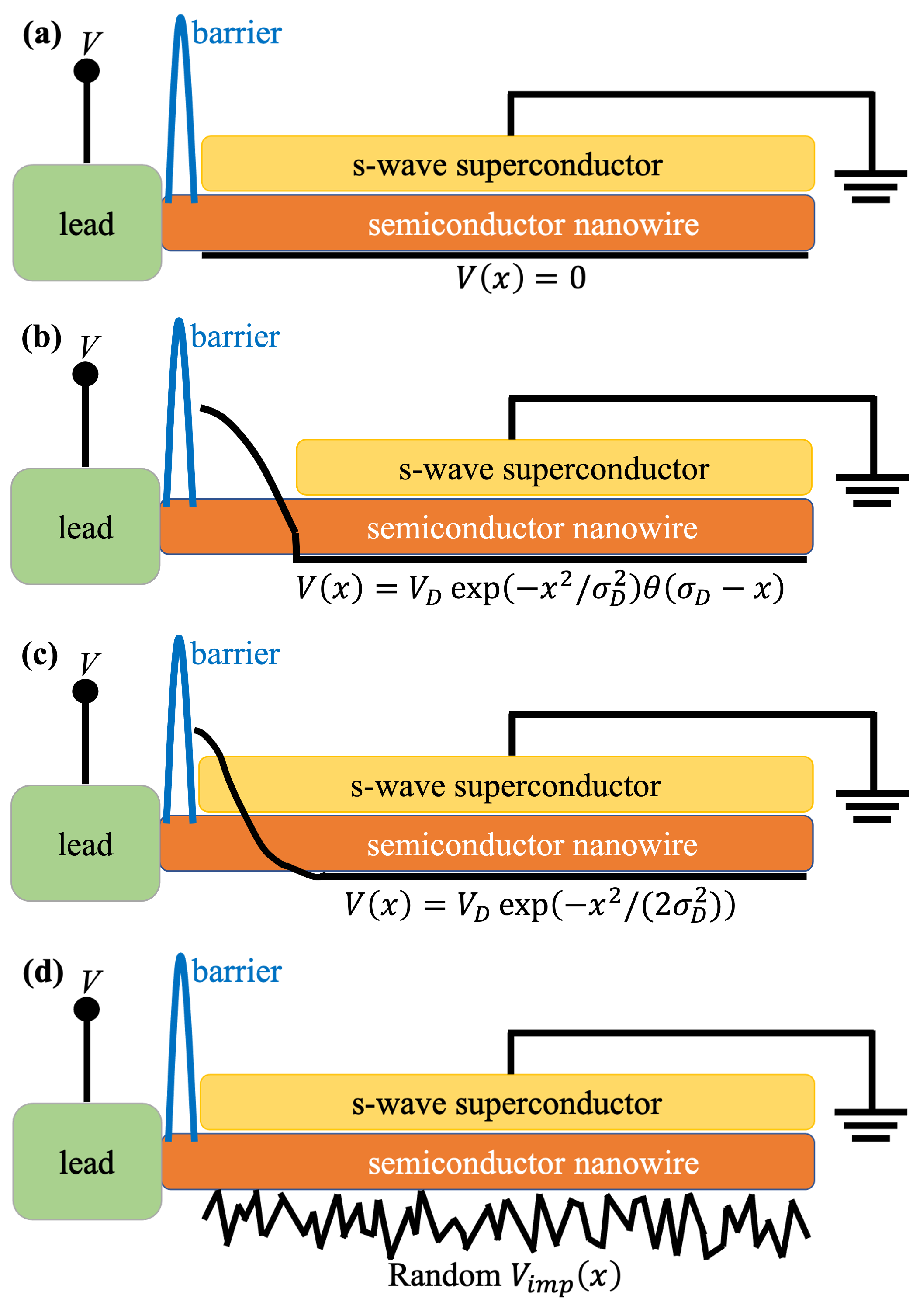}
	\caption{Schematic plots of the hybrid structures of a 1D superconductor-semiconductor nanowire and an attached lead, with different potentials. (a) Pristine nanowire: $V(x)=0$. (b) Nanowire with a SC-uncovered quantum dot $V(x)=V_D\exp(-x^2/\sigma_D^2)$ with length $\sigma_D$. (c) Nanowire with an inhomogeneous potential $V(x)=V_{D}\exp(-x^2/(2\sigma_D^2))$. (d) Nanowire with random disorder $V(x)$.}
	\label{fig:Scheme}
\end{figure}

In this section, we will describe the 1D SC-SM nanowire~\cite{Lutchyn2010Majorana,Sau2010Generic,Sau2010NonAbelian,Oreg2010Helical} as in Fig.~\ref{fig:Scheme}, which gives rise to ``good", ``bad", and ``ugly" ZBCPs, depending on the form of the potential in the real space~\cite{Pan2020Physical}. 
	
We use the minimal single-band model to describe the 1D superconductor-proximitized semiconductor nanowire with intrinsic Rashba spin-orbit coupling and external Zeeman-field-induced spin splitting, in the form of a Bogoliubov-de Gennes (BdG) Hamiltonian
\begin{equation}\label{hatH}
	\hat{H}=\frac{1}{2}\int_{0}^{L}dx\hat{\Psi}^\dagger(x)H_{\text{NW}}\hat{\Psi}(x)
\end{equation}
with
\begin{equation}\label{H_NW}
	\begin{aligned}
		H_{\text{NW}}(\omega)=&\left(-\frac{\hbar^2}{2m^*}\partial_x^2-i\alpha_R\partial_x\sigma_y-\mu+V(x)\right)\tau_z\\
		&+V_z\sigma_x+\Sigma(\omega,V_z)-i\Gamma,
	\end{aligned}
\end{equation}
where $m^*=0.015m_e$ is the effective mass of an electron ($m_e$ is the rest mass of an electron), $\mu$ is the chemical potential, and $\Gamma$ is an infinitesimal dissipation parameter included to avoid singularities in $G_0(E)$ from resonant transmission. The wave function in Eq.~\eqref{hatH} is $\hat{\Psi}(x)=\left(\hat{\psi}_\uparrow(x),\hat{\psi}_\downarrow(x),\hat{\psi}_\downarrow^\dagger(x),-\hat{\psi}_\uparrow^\dagger(x)\right)^T$ in Nambu space, with $\sigma_{x,y,z}(\tau_{x,y,z})$ being Pauli matrices in spin (particle-hole) space. The Rashba spin-orbit coupling with strength $\alpha_R$ is perpendicular to the wire~\cite{Bychkov1984Oscillatory} and the magnetic field $B$ is applied along the nanowire longitudinally such that the Zeeman term $V_z=\frac{1}{2}g\mu_B B$, where $\mu_B$ is Bohr magneton. $\Sigma(\omega,V_z)$ is a superconducting self-energy (see Appendix~\ref{sec:levelA_1} for details). We use the minimal model to simulate the nanowire instead of a more complex 3D model because it is found that the minimal model gives results very similar to complex models~\cite{Pan2021Quantized,Ahn2021Estimating}. In addition, a more complex model will have many more unknown parameters, making it less useful.
	
Besides the SC-SM nanowire itself, the normal lead attached to the end of the nanowire (as the left green block in Fig.~\ref{fig:Scheme}) is where the tunneling conductance is measured. A tunnel barrier induced at the interface of the normal-superconductor (NS) junction can be described by replacing $V(x)$ in Eq.~\eqref{H_NW} with a boxlike potential $V_{\text{barrier}}(x)$ (as the blue barrier potential at the left end of the nanowire in Fig.~\ref{fig:Scheme}) with the barrier potential height $E_{\text{barrier}}$ along with $\lambda(x)=0$, i.e., uncovered by the SC. Check Appendix~\ref{sec:levelA_3} for details.
	
We numerically compute the local differential tunneling conductance $G_0=dI_L/dV_L$ from the normal lead at the left end through the NS junction by the scattering matrix (S-matrix) method [Eq.~\eqref{G_sm}]. Specifically, we use Python package KWANT~\cite{Groth2014Kwant} to compute the differential conductance with the in-built scattering matrix derived from the known Hamiltonian. The conductance at zero temperature can be expressed and computed by the scattering matrix elements as follows: 
\begin{equation}\label{G_sm}
	G_0=N-\text{Tr}(r_{ee}r_{ee}^\dagger-r_{eh}r_{eh}^\dagger)
\end{equation}
in the unit of $e^2/h$, where $N$ is the number of the conducting channels in the lead, $r_{ee}$ is the normal reflection matrix, and $r_{eh}$ is the normal reflection matrix. In our system with only one-subband lead, $N=2$ counts the two spin modes. The conductance at finite temperature $G(V)$ can be further calculated by the convolution of the zero-temperature conductance $G_0(E)$ and the derivative of the Fermi-Dirac distribution $\partial f(E,T)/\partial E$, i.e.,
\begin{equation}\label{G_T}
	\begin{aligned}
		G(V)=&-\int_{-\infty}^{\infty}G_0(E)\frac{\partial f(E-V,T)}{\partial E}dE\\
		=&-\int_{-\infty}^{\infty} G_0(E)\left[\frac{1}{4T}\text{sech}^2\left(\frac{V-E}{2T}\right)\right]dE.
	\end{aligned}
\end{equation}
The above is the formalism for simulating the differential tunneling conductance in our results.
	
\subsection{Potential of Good/Bad/Ugly ZBCPs}\label{sec:level1_2_2}
The potential $V(x)$ in Eq.~\eqref{H_NW} determines whether the ZBCP belongs to the ``good", ``bad" or ``ugly" type~\cite{Pan2020Physical}. There are four kinds of potentials we set up for numerical simulations to discuss the quality factor of quantization for different types of ZBCPs. First of all, Eq.~\eqref{H_NW} displays as a pristine nanowire that can produce ``good" ZBCPs above TQPT field when $V(x)=0$, as in Fig.~\ref{fig:Scheme}(a). This Hamiltonian can definitely generates the genuine MZMs in the topological regime~\cite{Lutchyn2010Majorana,Sau2010Generic,Sau2010NonAbelian,Oreg2010Helical}. However, in the realistic experimental system, the unintentional potentials can produce ``bad" and ``ugly" ZBCPs.
	
The ``bad" ZBCPs are defined as those ZBCPs appearing in the topologically trivial regime, resulting from the deterministic spatially-varying potential. These ``bad'' ZBCPs are asscociated with so-called ABSs.~\cite{Kells2012Nearzeroenergy,Liu2012ZeroBias,Bagrets2012Class,Pikulin2012zerovoltage,Prada2012Transport,Sau2013Density,Mi2014Xshaped,Liu2017Andreev,Moore2018Quantized,Moore2018Twoterminal,Vuik2019Reproducinga} Such ABSs result from unintended quantum dots created when the lead is attached to the end of the nanowire. Such a quantum dot can arise either from mismatch of Fermi energies between the normal lead, semiconductor and the superconductor or from screened charged impurities or their combination~\cite{Liu2017Andreev,Moore2018Twoterminal}. In our model, the quantum dot hosts a Gaussian potential at the end of the nanowire, part of which is not covered by the parent SC, as in Fig.~\ref{fig:Scheme}(b). That is to say, the quantum dot potential is
\begin{equation}\label{V_QD}
	V(x)=V_D\exp\left(-\frac{x^2}{\sigma_D^2}\right)\theta(\sigma_D-x),
\end{equation}
where $V_D$ is the dot barrier height and $\sigma_D$ is the dot length. Also, the parent SC is mathematically expressed as 
\begin{equation}\label{SC_QD}
	\Delta_{\text{SC}}(x,V_z)=\Delta(V_z)\cdot\theta(x-\sigma_D),
\end{equation}
where $\Delta(V_z)$ has the same definition as Eq.~\eqref{SCgap}. The Heaviside step function $\theta(x)$ describes that the SC only covers the nanowire outside of the quantum dot. A variation of the quantum dot model with $\sigma'_D=\sqrt{2}\sigma_D$ to replace $\sigma_D$ in Eq.~\eqref{V_QD} can occur where the quantum dot potential extends into the superconductor as shown in Fig.~\ref{fig:Scheme}(c). This can be accomplished by dropping the factor $\Theta(x-\sigma_D)$ in both Eqs.~\eqref{V_QD} and~\eqref{SC_QD}. Results for the case of a ``bad'' potential that do not specify an $\sigma_D$ should be understood as being obtained from calculations which use this variant of the ``bad" potential.
	
The ``ugly" ZBCPs are defined as those ZBCPs showing up in the trivial regime, induced by random strong disorder. The typical potential that accounts for ``ugly" ZBCPs is the onsite disorder-induced random potential which follows an uncorrelated Gaussian distribution statistically with zero mean and standard deviation $\sigma_\mu$, i.e.,
\begin{equation}\label{V_imp}
	\begin{aligned}
		V(x)=&V_{\text{imp}}(x),\\
		\langle V_{\text{imp}}(x)\rangle=&0,\\
		\langle V_{\text{imp}}(x)V_{\text{imp}}(x')\rangle=&\sigma_\mu^2\delta(x-x'),
	\end{aligned}
\end{equation}
as in Fig.~\ref{fig:Scheme}(d). Each set of ``ugly" results in Sec.~\ref{sec:level1_3_3} is based on just one particular configuration of $V_{\text{imp}}(x)$, which is unpredictable contrary to the deterministic quantum dot potential in Eq.~\eqref{V_QD}. This random impurity potential can induce trivial ``ugly" ZBCPs which mimic ``good" ZBCPs.
	
\subsection{Wave functions}\label{sec:level1_2_3}
To determine the topological superconducting characteristic of a one dimensional system, it is necessary to look at the structure of the Majorana modes. Specifically, a topological superconductor is characterized by spatially well-separated Majorana wave functions~\cite{Huang2018Metamorphosis,Vuik2019Reproducinga,Stanescu2019Robust}. On the other hand, the ABSs display two overlapping wave functions at one end of the nanowire~\cite{Lai2019Presence}, even we construct their wave functions in the Majorana mode. The $4N$ component Nambu wave function $\Psi(x)$ for an $N$-site system, corresponding to any eigenenergy $\omega_0$ obtained from peaks of $\rho(\omega)$ in Eq.~\eqref{DOS} can be obtained as the eigenstate of $H_{\text{NW}}(\omega_0)$ with eigenvalue $\omega_0$. In a clean or weakly disordered system, one could determine the topological characteristic of the system by a straightforward calculation of the topological invariant~\cite{Kitaev2001Unpaired,Fulga2012Scattering,DasSarma2016How}. However, this approach is not well-defined for systems without a spectral or transmission gap, e.g., short strongly disordered nanowires. The systems of interest here, shown in Fig.~\ref{fig:Scheme}, are systems with a few sub-gap states. In this case, it is simpler to directly determine the topological character of the system by analyzing the low-energy wave functions directly as we describe below.
	
Majorana mode wave functions, even for a topological superconducting system of finite size, split into non-zero energy ($\omega_0=\pm\epsilon$) eigenstates $\Psi_\epsilon(x)$ and $\Psi_{-\epsilon}(x)$ that are not themselves Majorana (i.e., particle-hole symmetric). For a low-energy eigen-wave function $\Psi_\epsilon(x)$ corresponding to a positive energy $\epsilon$, one can use the particle-hole symmetry to define an orthogonal wave function $\Psi_{-\epsilon}(x)=\sigma_y\tau_y \Psi^*_\epsilon(x)$, which can be checked to be an eigenstate of $H_{\text{NW}}(-\epsilon)$ with eigenvalue $-\epsilon$. Then we can reconstruct the particle-hole symmetric Majorana wave functions as
\begin{equation}\label{WF_Maj}
	\begin{aligned}
		\Phi_A(x)=&\frac{1}{\sqrt{2}}\left[\Psi_\epsilon(x)+\Psi_{-\epsilon}(x)\right],\\
		\Phi_B(x)=&-\frac{i}{\sqrt{2}}\left[\Psi_\epsilon(x)-\Psi_{-\epsilon}(x)\right],
	\end{aligned}
\end{equation}
where $\Phi_{A,B}(x)=\left(\phi_\uparrow(x),\phi_\downarrow(x),\phi_\downarrow^*(x),-\phi_\uparrow^*(x)\right)^T$ are manifestly particle-hole symmetric. Note that this recipe suffers from a phase ambiguity of the eigenstate $\Psi_\epsilon(x)$ in the case of a general class D Hamiltonian, where it needs to be refined. This is not a problem for our case where the BdG Hamiltonian is real, which allows $\Psi_\epsilon(x)$ to be real. In general, $\Phi_{A,B}(x)$ are not the eigenfunctions of the BdG Hamiltonian, except when $\epsilon=0$, they represent the MZMs. However, when $\epsilon$ is much smaller than the superconducting gap, the off-diagonal matrix element of the Hamiltonian $H_{\text{NW}}(\omega=0)$ between $\Phi_{A,B}(x)$ is suppressed by a factor proportional to $\epsilon$. The matrix elements of $H_{\text{NW}}(\omega=0)$ between these states and any other excited states vanish as well. A system would be characterized as topological if the densities $|\Phi_\alpha(x)|^2$ corresponding to the state $\Phi_\alpha(x)$ are spatially separated.
	
If $|\Phi_A(x)|^2$ and $|\Phi_B(x)|^2$ are localized on both ends of the nanowire without overlapping, then we have a pair of MZMs. On the contrary, if $|\Phi_A(x)|^2$ and $|\Phi_B(x)|^2$ are clearly overlapped with each other, then the system hosts ABSs. When one of $|\Phi_{A,B}(x)|^2$ is localized on one end of the wire, while the other one is localized in the middle of the wire, then this pair can be quasi-Majorana bound states when they are partially overlapped with each other~\cite{Vuik2019Reproducinga,Stanescu2019Robust,Tian2021Distinguishing}.
	
\subsection{Quality factors}\label{sec:level1_2_4}
In this sub-section, we define the quality factors to quantify the stability of the quantized conductance plateau, which is the main focus of this paper. One of the most characteristic features of an MZM is quantized tunneling conductance~\cite{Sengupta2001Midgapa,Law2009Majorana,Flensberg2010Tunneling,Wimmer2011Quantum}. Specifically, the zero-bias (i.e., $V_{\text{bias}}=0$) conductance from a tunneling contact with one or two open channels into an MZM is predicted to be precisely quantized at $T=0$ for a sufficiently long topological wire even as the transmission of the tunnel contact becomes vanishingly small~\cite{Sengupta2001Midgapa,Law2009Majorana,Flensberg2010Tunneling,Wimmer2011Quantum}. This is counter-intuitive because the junction resistance, which can be estimated from the ``normal''-state conductance $G_N$, for $V_{\text{bias}}\gg \Delta$, diverges as the transmission of the tunnel contact is reduced. We use quotation marks over ``normal'' here to emphasize that this quantity is close to the actual normal-state conductance only in the limit that the superconducting gap is the smallest energy scale in the problem. However, since this is the most convenient quantity to measure,  we will refer to this quantity as the normal-state conductance in the rest of the draft. This large-series resistance should decrease the conductance of the MZM, which contradicts the theoretical quantization. Insight into this apparent counter-intuitive behavior is obtained by considering the analytic form for conductance into a Majorana $\gamma$, which is weakly coupled to the Majorana $\gamma'$ at the other end~\cite{Flensberg2010Tunneling}. One finds that the conductance is given by 
\begin{equation}\label{G0}
	G_0(V)=\frac{2e^2}{h}\frac{(2V\Gamma)^2}{(V^2-4t^2)^2+(2 V\Gamma)^2},
\end{equation}
where $t$ is the splitting of the Majorana modes and $\Gamma$ is the tunnel broadening~\cite{Flensberg2010Tunneling}, which vanishes with the normal-state conductance, i.e., $G_N\sim (2e^2/h)\Gamma/\Delta$. As the Majorana splitting $t$ vanishes, the above conductance takes the form of a Lorentzian with quantized height $(2e^2/h)$, but with a width $\Gamma$.  This result is changed at finite temperature according to Eq.~\eqref{G_T}, so the quantized height is reduced as $T\gtrsim\Gamma$ such that it approaches $G(V=0)\sim (2e^2/h)\Gamma/T$. The latter form is more consistent with an expectation of a conductance limited by the normal-state conductance $G_N$.
	
\begin{figure}
	\includegraphics[scale=0.6]{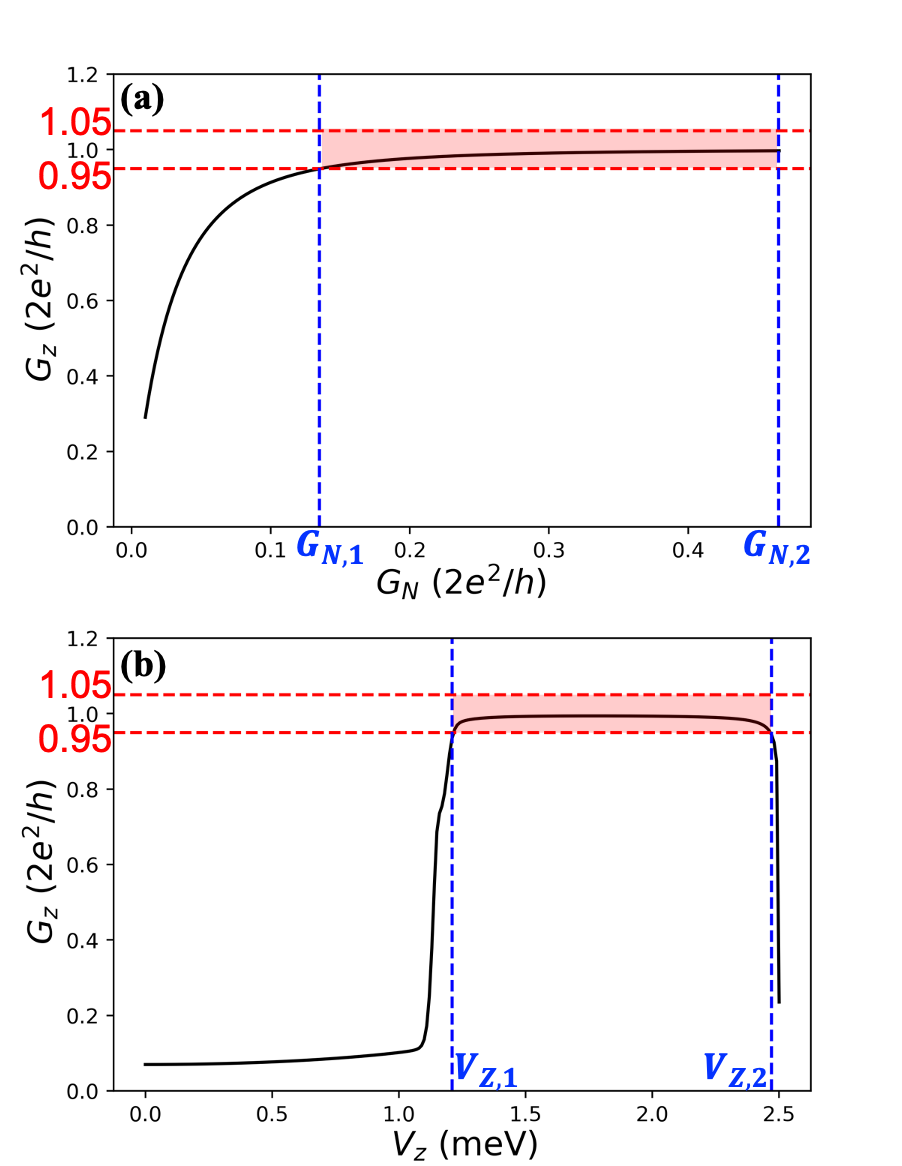}
	\caption{The tolerance factor for quantized conductance is $\epsilon=0.05$. (a) Black curve is the zero-bias conductance peak $G_z$ versus normal-metal conductance (above SC gap) $G_N$. The quality factor $F\equiv G_{N,2}/G_{N,1}$, where $G_{N,1}$ and $G_{N,2}$ are defined by the consecutive normal-metal range for which the corresponding zero-bias peak is quantized within $[1-\epsilon,1+\epsilon]$ in the unit of $2e^2/h$. (b) Black curve is the zero-bias conductance peak $G_z$ versus Zeeman field $V_z$. The quality factor $J\equiv V_{Z,2}/V_{Z,1}$, where $V_{Z,1}$ and $V_{Z,2}$ are defined by the consecutive Zeeman field range for which the corresponding zero-bias peak is quantized within $[1-\epsilon,1+\epsilon]$ in the unit of $2e^2/h$.}
	\label{fig:QualityFactor}
\end{figure}
	
The theoretically expected behavior for the zero-bias conductance into an MZM that was described in the last paragraph is confirmed by the conductance $G_z$ versus $G_N$ plot for an ideal Majorana wire (similar to that described in Sec.~\ref{sec:level1_2_2}) shown in Fig.~\ref{fig:QualityFactor}(a). The so-called normal-state conductance $G_N$ is ideally defined as the conductance without superconductivity. Since it is often non-trivial to remove superconductivity from a device without changing other factors, $G_N$ is practically measured (or calculated in theory) by taking an average of the conductance at $V_{\text{bias}}=\pm V_{\text{large}}$, where $V_{\text{large}}$ is a large bias voltage larger than the gap. It should be pointed out that $V_{\text{large}}$ should ideally be small compared to the Fermi energy, so as not to affect the transmission significantly. Unfortunately, the difference in the conductance at positive and negative voltages confirms that this critierion is often not satisfied leading to some (but not significant) ambiguity in $G_N$. The variation of $G_N$ in Fig.~\ref{fig:QualityFactor}(a) is obtained by varying the barrier height, which in the experiment can be done by tuning a tunnel gate voltage~\cite{Zhang2018Quantized,Zhang2021Large} in semiconductor setups or changing the tip-sample distance in scanning tunneling microscopy (STM)~\cite{Zhu2020Nearly}. Because of the finite temperature used in the calculation, the conductance $G_z$ shown in Fig.~\ref{fig:QualityFactor}(a) shows a nearly quantized plateau at high values of the normal-state conductance $G_N$, which then decreases to zero linearly as $G_N$ is reduced to zero as expected based on the fundamental theory discussed in the last paragraph.

The $G_N$-independent quantized plateau is a signature of MZMs. While the conductance into a superconductor without an MZM can be tuned to quantization by varying $G_N$ and other parameters, we do not expect the quantization to be robust. However, the conductance into an MZM is not precisely quantized because of finite-temperature and finite-size effects. In this paper, we propose to distinguish MZMs from other superconducting bound states by quantifying the robustness of the conductance plateau seen in Fig.~\ref{fig:QualityFactor}(a). To do this, we first identify the largest continuous interval [$G_{N,1}$, $G_{N,2}$] on the $x$ axis over which the zero-bias conductance $G_z$ is within a tolerance $\epsilon$ of quantization, i.e., $|G_z-1|<\epsilon$ (in units of $2e^2/h$). We then assign a quality factor
\begin{equation}\label{F}
	F=\frac{G_{N,2}}{G_{N,1}}
\end{equation}
as a degree to which the conductance into the MZM is quantized. In Fig.~\ref{fig:QualityFactor}(a), we set $\epsilon=0.05$ as an example, meaning as long as the ZBCP is above $95\%$ of $2e^2/h$ and below $105\%$ of $2e^2/h$ [within pink region in Fig.~\ref{fig:QualityFactor}(a)], then we take the ZBCP as a ``well-quantized" ZBCP. In this paper, we also demonstrate the numerical results for $\epsilon=0.10$ and $\epsilon=0.20$ for comparison. In some special cases, where the ZBCP over the visible range is sectioned into several parts [e.g., Fig.~\ref{fig:BadZBP_10}(f)], we take the largest consecutive normal-metal range to define $G_{N,1}$ and $G_{N,2}$.
	
The conductance into an MZM should be similar to other parameters as well because of the robustness of the topological phase. The Zeeman splitting $V_z$ that is controlled by tuning the applied magnetic field is one such parameter. While there is no fundamental bound on the extent of the topological phase in $V_z$, an MZM system associated with a reasonably large topological gap is expected to be robust to changes of $V_z$ as long as the topological gap is not destroyed. Here, we will quantify the robustness of the MZM to changes in the Zeeman field using a quality factor $J$, which is defined in an analogous way to the tunnel gate quality factor $F$ defined in the last paragraph. Figure~\ref{fig:QualityFactor}(b) is the plot of ZBCP $G_z$ versus Zeeman field. This is the conductance linecut as in Fig.~\ref{fig:GoodZBP_2}(e), which can be extracted from Fig.~\ref{fig:GoodZBP_2}(a) with only $V_{\text{bias}}=0$. The definition of quality factor $J$ is similar to $F$ as illustrated in the previous paragraph, except that we change the normal-metal conductance $G_N$ part to Zeeman field $V_z$ for the quality factor $J$. Formally, it is defined as $J\equiv V_{Z,2}/V_{Z,1}$, where $[V_{Z,1},V_{Z,2}]$ is the range over which $\left|G_z-1\right|<\epsilon$ in the unit of $2e^2/h$ and the tolerance factor for quantized conductance $\epsilon$ is a small number. Same as Fig.~\ref{fig:QualityFactor}(a), we set $\epsilon=0.05$, $5\%$ difference from quantized value $2e^2/h$ as highlighted in the pink region in Fig.~\ref{fig:QualityFactor}(b), as an example. All the explanations for $F$ can also be applied to $J$ as long as we replace $G_{N,1}$ and $G_{N,2}$ by $V_{Z,1}$ and $V_{Z,2}$, respectively.
	
The definitions of $F$ and $J$ as presented so far in this sub-section are not defined in cases where the zero-bias conductance $G_z$ does not cross the quantized values. In this case, we formally define $F$ and $J$ to be zero. In non-topological cases, where the zero-bias conductance $G_z$ crosses the quantized value, $F$ and $J$ are also slightly higher but close to 1. In contrast, we will see from our numerical results in Sec.~\ref{sec:level1_3} that $F$ and $J$ can be much larger than 1 in the ideal case of low temperature toplogical superconductors. The main goal of our paper is to study $F$ and $J$ for various models to determine if they can be used to distinguish MZMs from other non-topological sources of ZBCPs. 
	
\section{Results}\label{sec:level1_3}
In this section, we present conductance plots for various tunnel gate strengths and magnetic field strengths to determine the robustness of the ZBCP quantization to the variety of perturbations. The finite-temperature conductance plots will be used to determine the quality factors $F$ and $J$ for the ZBCP for a variety of models. The quality factors $F$ and $J$ for the ideal MZM will turn out to be limited by temperature, topological gap and length. The goal of this section will be to compare this quality factor for the so-called ``good'' ZBCP (i.e., ideal MZM) with other non-topological models for the ZBCP (i.e., the ``bad'' and ``ugly'' ZBCPs discussed in Sec.~\ref{sec:level1_2_2}). Since the quantities plotted in this section are the typical ones that are measured in most MZM experiments~\cite{Nichele2017Scaling,Gul2018Ballistic,Zhang2018Quantized,Zhu2020Nearly,Zhang2021Large}, we believe that the conclusions about $F$ and $J$ obtained from these results can directly be applied to experiments on MZMs, as we will discuss in Sec.~\ref{sec:level1_4}. 
	
\subsection{``Good" ZBCP}\label{sec:level1_3_1} % one set
\begin{figure*} 
	\includegraphics[scale=0.42]{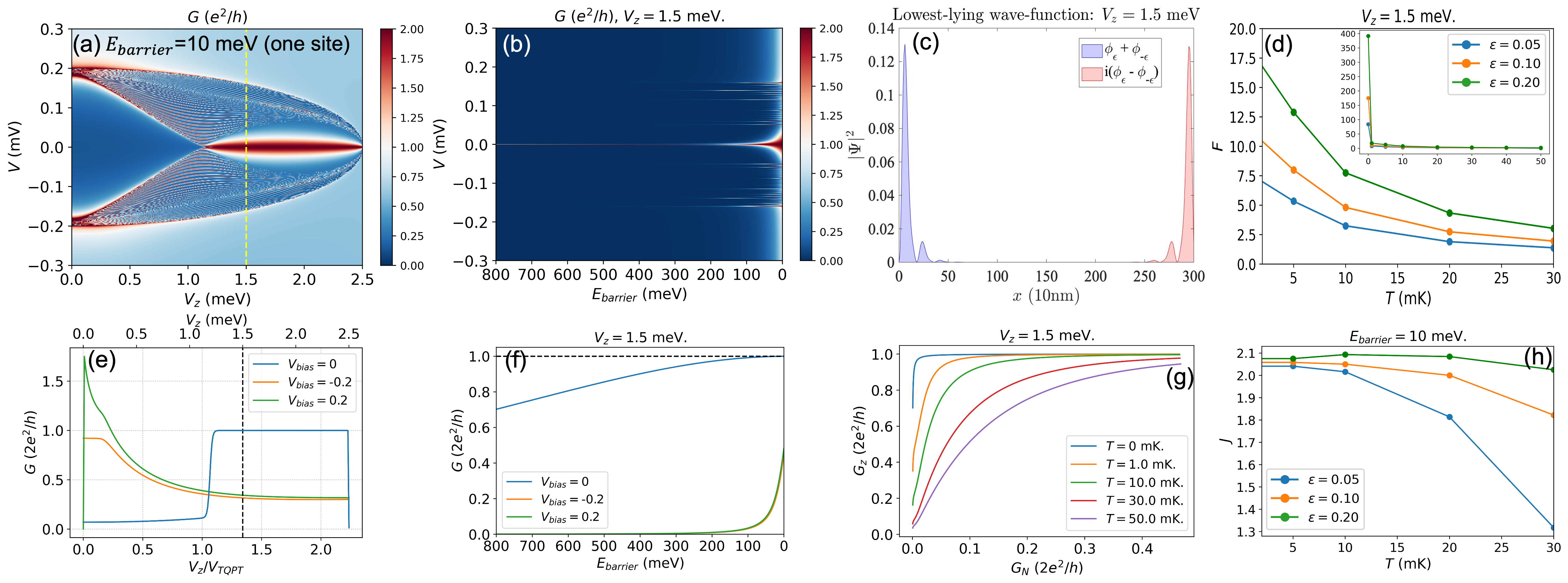}
	\caption{Numerical results for the ``good" ZBCP. Parameters: $\alpha=3.0$ meV, $\Delta_0=0.2$ meV, $V_c=2.5$ meV, $L=3.0$ $\mu$m, $\mu=0.5$ meV, $\lambda=1.0$ meV, and $\gamma=10^{-4}$ meV. The TQPT field is at $V_{\text{TQPT}}=\sqrt{\lambda^2+\mu^2}=1.118$ meV. (a) Conductance false-color plot as a function of bias voltage $V$ and Zeeman field $V_z$, with the fixed tunneling barrier height at $E_{\text{barrier}}=10$ meV. (b) Conductance false-color plot as a function of bias voltage $V$ and tunneling barrier height $E_{\text{barrier}}$, with the fixed Zeeman field at $V_z=1.5$ meV. (c) Lowest-lying wave-function probability density $|\Psi|^2$ as a function of nanowire position $x$, with fixed $V_z=1.5$ meV and $E_{\text{barrier}}=10$ meV. (d) Quality factor $F$ as a function of temperature $T$ for three different tolerance factors $\epsilon$, with the fixed $V_z=1.5$ meV. The inset figure gives an overall trend starting from $T=0$. At $T=10$ mK, $F=3.25$ for $\epsilon=0.05$. At $T=20$ mK, $F=2.74$ for $\epsilon=0.1$. (e) Conductance line cuts as a function of Zeeman field $V_z$ for three different bias voltages $V_{\text{bias}}$, with the fixed $E_{\text{barrier}}=10$ meV. The black dashed line marks $V_z=1.5$ meV. (f) Conductance line cuts as a function of tunneling barrier height $E_{\text{barrier}}$ for three different bias voltages $V_{\text{bias}}$, with the fixed $V_z=1.5$ meV. (g) Zero-bias conductance $G_z$ as a function of normal-metal conductance $G_N$ for five different temperatures $T$, with the fixed $V_z=1.5$ meV. (h) Quality factor $J$ as a function of temperature $T$ for three different tolerance factors $\epsilon$, with the fixed $E_{\text{barrier}}=10$ meV.}
	\label{fig:GoodZBP_2}
\end{figure*}
	
ZBCPs associated with MZMs at the ends of topological superconducting nanowires are referred to as ``good" ZBCPs~\cite{Pan2020Physical}. Theoretically, they can be produced above the TQPT field in the simple pristine SC-SM nanowire model as Fig.~\ref{fig:Scheme}(a), which is the setup for the numerical results in Fig.~\ref{fig:GoodZBP_2}. While this model is not particularly relevant for experiment and one expects at least weak versions of the effects shown in Figs.~\ref{fig:Scheme}(b)-(d) to appear in reality, the results in this sub-section  will serve as a reference for the signatures of a topological superconductor.
	
Figure~\ref{fig:GoodZBP_2}(a) shows the conductance (at zero temperature) as a false-color plot versus the bias voltage $V$ and Zeeman field $V_z$. As expected from previous works~\cite{Sengupta2001Midgapa,Law2009Majorana,Flensberg2010Tunneling,Wimmer2011Quantum}, we observe that a nearly quantized ZBCP appears at a Zeeman field above the TQPT field $V_z>V_{\text{TQPT}}=1.118$ meV. The topologically non-trivial origin of this ZBCP can be seen from Fig.~\ref{fig:GoodZBP_2}(c), which shows the lowest-lying wave-function probabilities $|\Phi_{A,B}(x)|^2$ (see Sec.~\ref{sec:level1_2_3}) at $V_z=1.5$ meV$>V_{\text{TQPT}}$ [yellow dashed line in Fig.~\ref{fig:GoodZBP_2}(a)]. The localization of $|\Phi_{A,B}(x)|^2$ at opposite ends of the wire confirms that the ZBCP at $V_z=1.5$ meV arises from topological MZM modes. Unfortunately, the spatial structure of the wave function is not accessible in a transport experiment in a nanowire. On the other hand, as discussed in the Introduction and Sec.~\ref{sec:level1_2_4}, the quantization of the ZBCP associated with a topological MZM should be robust to changes in the barrier height. This is consistent with the results shown in Fig.~\ref{fig:GoodZBP_2}(b), which shows the conductance false-color plot as a function of bias voltage $V$ and tunneling barrier height $E_{\text{barrier}}$ at the fixed Zeeman field $V_z=1.5$ meV. The plot shows that the height of the ZBCP associated with the MZM is largely unchanged with increasing tunnel barrier height $E_{\text{barrier}}$, while the width of the ZBCP decreases. This is further confirmed by the linecuts from Fig.~\ref{fig:GoodZBP_2}(b) at $V_{\text{bias}}=0$ that are shown in Fig.~\ref{fig:GoodZBP_2}(f), where the ZBCP height, $G_z$, is found to change by less than $5\%$ from quantization. Figure~\ref{fig:GoodZBP_2}(f) also shows line cuts from Fig.~\ref{fig:GoodZBP_2}(b) at $V_{\text{bias}}=\pm 0.2$ meV. In contrast to the ZBCP height, these linecuts, which may be interpreted as the normal-state conductance, change significantly with the tunnel barrier height $E_{\text{barrier}}$. In fact, the tunnel barrier height $E_{\text{barrier}}$ is not directly comparable to anything measurable in experiments since tunnel gates have so-called lever arms that are determined by complicated capacitance structures. On the other hand, the normal-state conductance $G_N$ determined by the plots from Fig.~\ref{fig:GoodZBP_2}(f) at $V_{\text{bias}}=\pm 0.2$ meV provides a way to quantify the tunnel barrier height in a way that is directly comparable to experiments. One subtlety that arises in semiconductor systems is that the conductance at the non-zero biases in Fig.~\ref{fig:GoodZBP_2}(f) can be different in semiconductor systems, where the Fermi energy could be comparable to the superconducting gap $\Delta\sim 0.2$ meV. This is remedied by defining $G_N$ to be the average between the conductances at $V_{\text{bias}}=\pm 0.2$ meV.
	
Thus, to obtain a result that can be directly compared to experiments, we plot $G_z$ versus $G_N$ in Fig.~\ref{fig:GoodZBP_2}(g), while keeping $V_z=1.5$ meV, so the wire is in the topological superconducting phase [yellow dashed line in Fig.~\ref{fig:GoodZBP_2}(a)]. Consistent with our discussion in the previous paragraph about Fig.~\ref{fig:GoodZBP_2}(f), we find that the ZBCP height does not change significantly as $G_N$ is reduced at temperature $T=0$. This is in contrast to the results at finite temperature $T>0$ that are obtained using Eq.~\eqref{G_T} and show that $G_z$ goes to zero as $G_N$ decreases, as expected from the discussion in Sec.~\ref{sec:level1_2_4}. The results in Fig.~\ref{fig:GoodZBP_2}(g) make it clear that the robustness of the quantized ZBCP to changing the tunnel barrier, even in the case of ideal MZMs, would be limited by the temperature at which the measurement is performed. The degree of robustness of the ZBCP quantization can be characterized by the quality factor $F$ that was defined in Sec.~\ref{sec:level1_2_4} as the relative size of the range of $G_N$ where the conductance $G_z$ is within a tolerance $\epsilon$ from quantization. The plot of the quality factor $F$ versus temperature shown in Fig.~\ref{fig:GoodZBP_2}(d), shows that while the quality factor $F$ associated with MZMs can be quite large (i.e., more than 80 at $T=0$), $F$ decreases quite rapidly as the temperature $T$ becomes comparable to the topological superconducting gap. The main figure in Fig.~\ref{fig:GoodZBP_2}(d) is to show the detailed variations of $F$ in the experimental temperature range from $T=2$ mK to $T=30$ mK. The inset figure starting from $T=0$ gives an overall trend to show how sharply $F$ decreases from zero temperature to finite temperature. Since we only want to demonstrate how sharp the change of $F$ is by the temperature variations, the tick numbers in the inset figure are not the key points here. While the quantitative value of $F$ increases with the increasing tolerance factor $\epsilon$ at which the quality factor is calculated, the qualitative behavior does not appear to be significantly affected by the value of $\epsilon$.
	
Ideal MZMs are expected to arise in a topological superconducting phase that should be robust to variations in the parameter such as the Zeeman potential controlled by the applied magnetic field. This is consistent with the result for the zero-bias conductance seen in Fig.~\ref{fig:GoodZBP_2}(e) where we find that the zero-bias (i.e., $V_{\text{bias}}=0$) conductance starts out small in the non-topological regime (i.e., $V_z<V_{\text{TQPT}}$), but then sharply rises to a quantized plateau for $V_z>V_{\text{TQPT}}$. Similar to the robustness to the tunnel barrier height seen in Fig.~\ref{fig:GoodZBP_2}(f), the result in Fig.~\ref{fig:GoodZBP_2}(e) shows that the ZBCP associated with a topological MZM is expected to be quite robust to variations in the Zeeman potential. Following the discussion in Sec.~\ref{sec:level1_2_4}, this robustness can be quantified through defining a parameter $J$, which is plotted as a function of temperature in Fig.~\ref{fig:GoodZBP_2}(h). As with the case of the quality factor $F$ associated with tunnel gate robustness, the Zeeman-field quality factor $J$ also decreases with temperature in a way that is qualitatively independent of the tolerance $\epsilon$ used in the definition of $J$.
	
\subsection{``Bad" ZBCP}\label{sec:level1_3_2} % two sets
\begin{figure*} % "Bad" bad
	\includegraphics[scale=0.41]{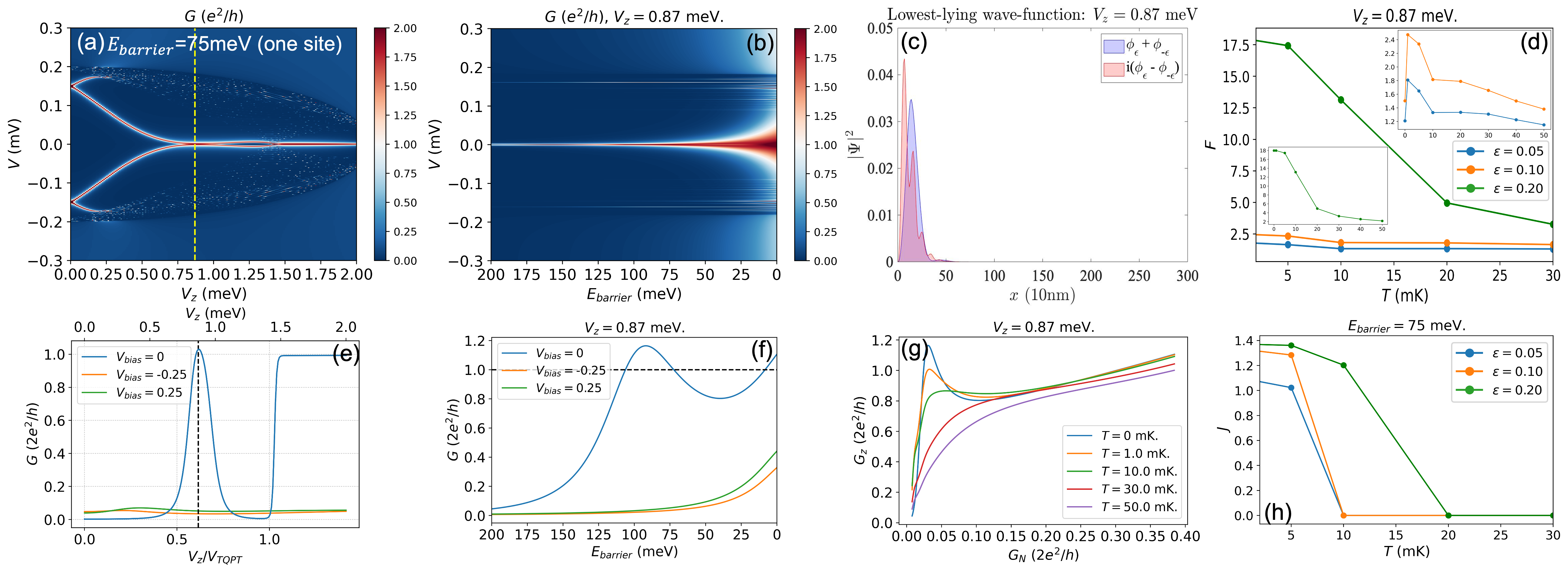}
	\caption{Numerical results for the ``bad" ZBCP. Parameters: $\alpha=2.51$ meV, $\Delta_0=0.2$ meV, $V_c=2.1$ meV, $L=3.0$ $\mu$m, $\mu=1.0$ meV, $\lambda=1.0$ meV, and $\gamma=10^{-4}$ meV. The TQPT field is at $V_{\text{TQPT}}=\sqrt{\lambda^2+\mu^2}=1.414$ meV. The parameters for the quantum dot on the left end are: $V_D=0.3$ meV and $\sigma_D=0.15$ $\mu$m. (a) Conductance false-color plot as a function of bias voltage $V$ and Zeeman field $V_z$, with the fixed tunneling barrier height at $E_{\text{barrier}}=75$ meV. (b) Conductance false-color plot as a function of bias voltage $V$ and tunneling barrier height $E_{\text{barrier}}$, with the fixed Zeeman field at $V_z=0.87$ meV. (c) Lowest-lying wave-function probability density $|\Psi|^2$ as a function of nanowire position $x$, with fixed $V_z=0.87$ meV and $E_{\text{barrier}}=75$ meV. (d) Quality factor $F$ as a function of temperature $T$ for three different tolerance factors $\epsilon$, with the fixed $V_z=0.87$ meV. The inset figure gives an overall trend starting from $T=0$. At $T=10$ mK, $F=1.331$ for $\epsilon=0.05$. At $T=20$ mK, $F=1.789$ for $\epsilon=0.1$. (e) Conductance line cuts as a function of Zeeman field $V_z$ for three different bias voltages $V_{\text{bias}}$, with the fixed $E_{\text{barrier}}=75$ meV. The black dashed line marks $V_z=0.87$ meV. (f) Conductance line cuts as a function of tunneling barrier height $E_{\text{barrier}}$ for three different bias voltages $V_{\text{bias}}$, with the fixed $V_z=0.87$ meV. (g) Zero-bias conductance $G_z$ as a function of normal-metal conductance $G_N$ for five different temperatures $T$, with the fixed $V_z=0.87$ meV. (h) Quality factor $J$ as a function of temperature $T$ for three different tolerance factors $\epsilon$, with the fixed $E_{\text{barrier}}=75$ meV.}
	\label{fig:BadZBP_10}
\end{figure*}
	
\begin{figure*} % "Good" bad
	\includegraphics[scale=0.41]{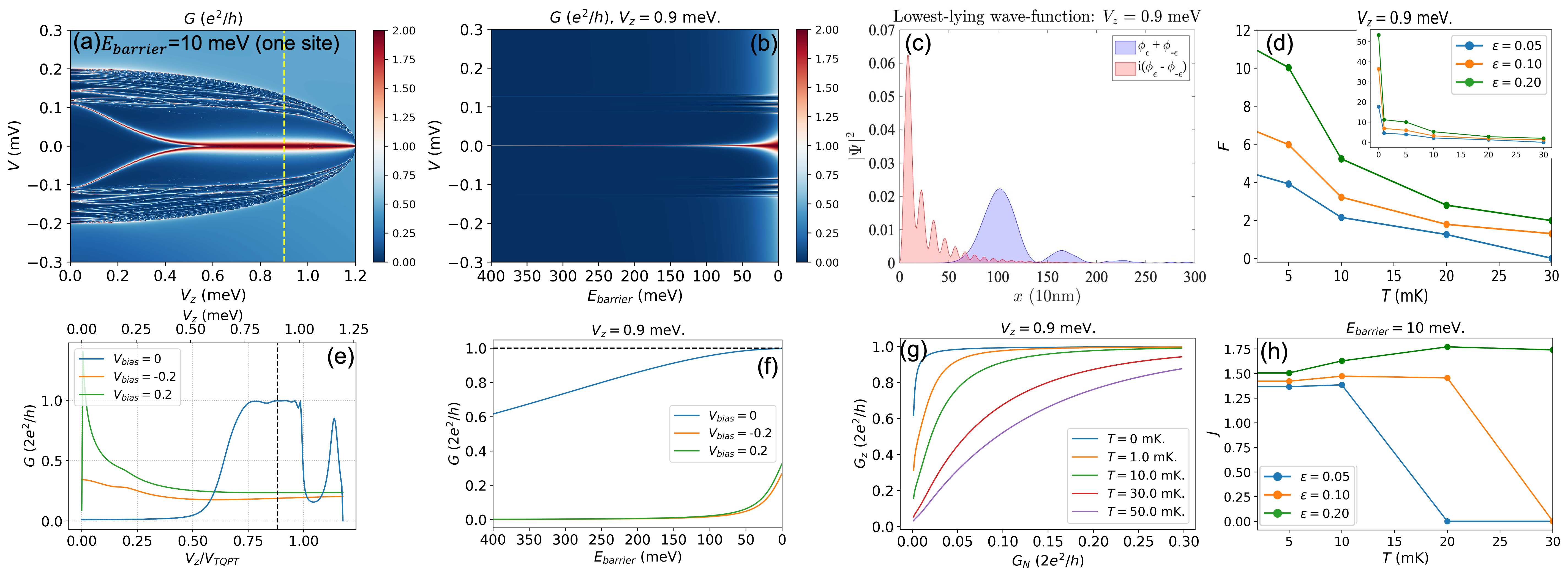} 
	\caption{Numerical results for the ``bad" ZBCP. Parameters: $\alpha=2.5$ meV, $\Delta_0=0.2$ meV, $V_c=1.2$ meV, $L=3.0$ $\mu$m, $\mu=1.0$ meV, $\lambda=0.2$ meV, and $\gamma=10^{-4}$ meV. The TQPT field is at $V_{\text{TQPT}}=\sqrt{\lambda^2+\mu^2}=1.020$ meV. The parameters for the inhomogeneous potential on the left end are: $V_D=1.2$ meV and $\sigma_D=0.4$ $\mu$m ($\sigma'_D=\sqrt{2}\times 0.4$ $\mu$m). (a) Conductance false-color plot as a function of bias voltage $V$ and Zeeman field $V_z$, with the fixed tunneling barrier height at $E_{\text{barrier}}=10$ meV. (b) Conductance false-color plot as a function of bias voltage $V$ and tunneling barrier height $E_{\text{barrier}}$, with the fixed Zeeman field at $V_z=0.9$ meV. (c) Lowest-lying wave-function probability density $|\Psi|^2$ as a function of nanowire position $x$, with fixed $V_z=0.9$ meV and $E_{\text{barrier}}=10$ meV. (d) Quality factor $F$ as a function of temperature $T$ for three different tolerance factors $\epsilon$, with the fixed $V_z=0.9$ meV. The inset figure gives an overall trend starting from $T=0$. At $T=10$ mK, $F=2.148$ for $\epsilon=0.05$. At $T=20$ mK, $F=1.785$ for $\epsilon=0.1$. (e) Conductance line cuts as a function of Zeeman field $V_z$ for three different bias voltages $V_{\text{bias}}$, with the fixed $E_{\text{barrier}}=10$ meV. The black dashed line marks $V_z=0.9$ meV. (f) Conductance line cuts as a function of tunneling barrier height $E_{\text{barrier}}$ for three different bias voltages $V_{\text{bias}}$, with the fixed $V_z=0.9$ meV. (g) Zero-bias conductance $G_z$ as a function of normal-metal conductance $G_N$ for five different temperatures $T$, with the fixed $V_z=0.9$ meV. (h) Quality factor $J$ as a function of temperature $T$ for three different tolerance factors $\epsilon$, with the fixed $E_{\text{barrier}}=10$ meV.}
	\label{fig:BadZBP_11}
\end{figure*}
	
Let us now consider the quality factor of so-called ``bad" ZBCPs, which arise as a result of an inhomogeneous potential near the end of the nanowire as shown in Figs.~\ref{fig:Scheme}(b) and ~\ref{fig:Scheme}(c). It has been shown previously that such an inhomogeneous potential can lead to ZBCPs~\cite{Liu2017Andreev}, which can have nearly quantized conductance~\cite{Vuik2019Reproducinga}. Therefore, it is interesting to understand if the robustness of the quantization is able to distinguish between such ABSs that give rise to ``bad'' ZBCPs and the topological MZMs discussed in the previous sub-section. 
	
The conductance color plot in Fig.~\ref{fig:BadZBP_10}(a), based on the setup in Fig.~\ref{fig:Scheme}(b), clearly shows a pair of conductance peaks at finite energy merge together into a ZBCP that is qualitatively quite similar to the topological result shown in Fig.~\ref{fig:GoodZBP_2}(a). Considering the zero-bias conductance line cut shown in Fig.~\ref{fig:BadZBP_10}(e), we observe that unlike the ideal MZM case, the zero-bias conductance below the TQPT at $V_z=0.87$ meV [marked by the black dashed line in Fig.~\ref{fig:BadZBP_10}(e) or the yellow dashed line in Fig.~\ref{fig:BadZBP_10}(a)] is near the quantized result $2e^2/h$ predicted for the topological case. A closer examination of the ZBCP in Fig.~\ref{fig:BadZBP_10}(a) suggests that the reduction of the ZBCP seen in Fig.~\ref{fig:BadZBP_10}(e) is likely a result of splitting of the ZBCP for $V_z>0.87$ meV. To study if the tunnel gate robustness of MZMs seen in the last sub-section applies to this nearly quantized ZBCP, we study the ZBCP height with varying tunnel barrier height. As seen from  Fig.~\ref{fig:BadZBP_10}(b), we see that the ZBCP height appears to remain nearly constant as the width of the ZBCP changes, quite similar to the case of ideal MZMs. However, a closer examination of the quantization using the $V_{\text{bias}}=0$ line cut shown in Fig.~\ref{fig:BadZBP_10}(f), shows that the ZBCP height at $V_z=0.87$ meV varies by a substantial amount as the tunnel barrier is varied. This variation can also be observed by considering the ZBCP height as a function of the normal-state conductance shown in Fig.~\ref{fig:BadZBP_10}(g). In contrast to the case of the ideal MZM discussed in the last paragraph, we find that the ZBCP height overshoots the quantized value at low temperatures and small normal-state conductance $G_N$. This is consistent with the relatively small value of the quality factor $F$ compared to the topological case as seen in Fig.~\ref{fig:BadZBP_10}(d). The quality factor $F$ for $\epsilon=0.05$ at $10$ mK in this case is below $1.5$. Interestingly, the quality factor $F$ can become quite large if we choose a tolerance of  $\epsilon=0.2$, which can be expected from the fact that the conductance $G_z$ in Fig.~\ref{fig:BadZBP_10}(g) remains somewhat close to quantized. These observations together with the weak splitting of the ZBCP seen in Fig.~\ref{fig:BadZBP_10}(a) can be understood in terms of the Majorana decompositions of the low-energy wave functions shown in Fig.~\ref{fig:BadZBP_10}(c). These wave functions show clearly that the system is non-topological because both Majorana components are strongly overlapping at the same end. However, one of the components has a stronger spatial modulation, suggesting having a strong weight at a different Fermi point. This suppresses the overlap between the two states and ensures that only one of the states can couple strongly to the lead, explaining the observations made about Figs.~\ref{fig:BadZBP_10}(a) and ~\ref{fig:BadZBP_10}(g). Because of the splitting of the ZBCP as a function of Zeeman field seen in Fig.~\ref{fig:BadZBP_10}(a), the ZBCP quantization in Fig.~\ref{fig:BadZBP_10}(e) is found to survive over a rather narrow range of Zeeman potential $V_z$. This leads to the suppressed value of the quality factor $J$ seen in Fig.~\ref{fig:BadZBP_10}(h) relative to the topological value.
	
Figure~\ref{fig:Scheme}(c) shows another inhomogeneous potential configuration that leads to the conductance shown in Fig.~\ref{fig:BadZBP_11} with a bad ZBCP. While the conductance peaks shown in Fig.~\ref{fig:BadZBP_11}(a) is qualitatively similar to the ``bad'' ZBCP in Fig.~\ref{fig:BadZBP_10}(a) that we discussed so far, the quantization of the ZBCP at $V_z=0.9$ meV, based on Figs.~\ref{fig:BadZBP_11}(b) and ~\ref{fig:BadZBP_11}(d)-~\ref{fig:BadZBP_11}(g) appears more robust relative to Fig.~\ref{fig:BadZBP_10}.  In fact, the quality factor $F$ in Fig.~\ref{fig:BadZBP_11}(d) associated with the ZBCP at temperature of 20 mK and $\epsilon=0.1$ is $1.79$, which is closer to the ideal value of $2.74$ in Fig.~\ref{fig:GoodZBP_2}(d) relative to the previous ``bad'' ZBCP [Fig.~\ref{fig:BadZBP_10}(d)]. As will be discussed in more detail later, this can be understood from the fact that the Majorana decomposition [Fig.~\ref{fig:BadZBP_11}(c)] shows a pair of partially separated modes, which have been described as quasi-Majoranas~\cite{Vuik2019Reproducinga}. Such separated segments between spatially separated MZMs may be thought of as topological in their own right to the extent that the overlap may be ignored and the lead (at the left end in this case) is only coupled to one Majorana mode but not the other. This explains the high value of $F$ quality factor in this case of ``bad'' ZBCP. The $J$ quality factor shown in Fig.~\ref{fig:BadZBP_11}(h) also turns out to be intermediate between the ideal case in Fig.~\ref{fig:GoodZBP_2}(h) and the previous bad ZBCP shown in Fig.~\ref{fig:BadZBP_10}(h).
	
\subsection{``Ugly" ZBCP}\label{sec:level1_3_3} % three sets
\begin{figure*} % "Bad" ugly
	\includegraphics[scale=0.41]{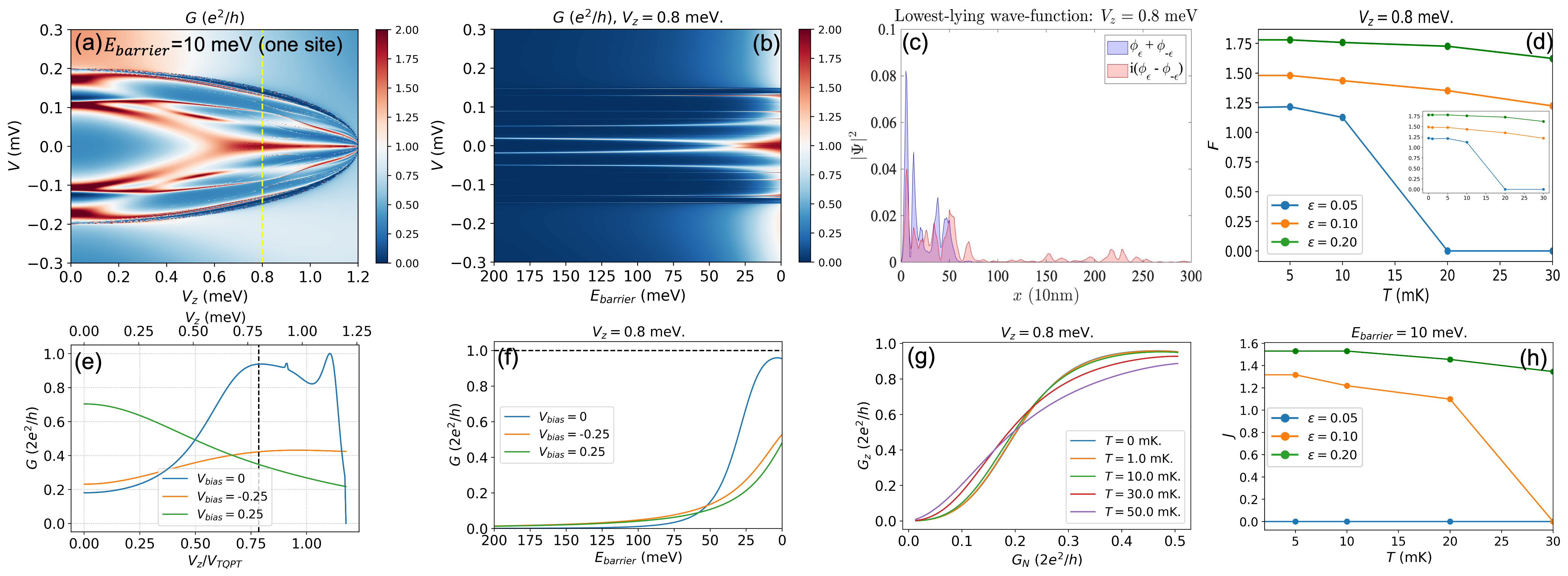}
	\caption{Numerical results for the ``ugly" ZBCP. Parameters: $\alpha=2.5$ meV, $\Delta_0=0.2$ meV, $V_c=1.2$ meV, $L=3.0$ $\mu$m, $\mu=1.0$ meV, $\lambda=0.2$ meV, and $\gamma=10^{-4}$ meV. The TQPT field is at $V_{\text{TQPT}}=\sqrt{\lambda^2+\mu^2}=1.020$ meV. The parameter for the on-site random potential is: $\sigma_\mu=2.0$ meV. (a) Conductance false-color plot as a function of bias voltage $V$ and Zeeman field $V_z$, with the fixed tunneling barrier height at $E_{\text{barrier}}=10$ meV. (b) Conductance false-color plot as a function of bias voltage $V$ and tunneling barrier height $E_{\text{barrier}}$, with the fixed Zeeman field at $V_z=0.8$ meV. (c) Lowest-lying wave-function probability density $|\Psi|^2$ as a function of nanowire position $x$, with fixed $V_z=0.8$ meV and $E_{\text{barrier}}=10$ meV. (d) Quality factor $F$ as a function of temperature $T$ for three different tolerance factors $\epsilon$, with the fixed $V_z=0.8$ meV. The inset figure gives an overall trend starting from $T=0$. At $T=10$ mK, $F=1.126$ for $\epsilon=0.05$. At $T=20$ mK, $F=1.352$ for $\epsilon=0.1$. (e) Conductance line cuts as a function of Zeeman field $V_z$ for three different bias voltages $V_{\text{bias}}$, with the fixed $E_{\text{barrier}}=10$ meV. The black dashed line marks $V_z=0.8$ meV. (f) Conductance line cuts as a function of tunneling barrier height $E_{\text{barrier}}$ for three different bias voltages $V_{\text{bias}}$, with the fixed $V_z=0.8$ meV. (g) Zero-bias conductance $G_z$ as a function of normal-metal conductance $G_N$ for five different temperatures $T$, with the fixed $V_z=0.8$ meV. (h) Quality factor $J$ as a function of temperature $T$ for three different tolerance factors $\epsilon$, with the fixed $E_{\text{barrier}}=10$ meV.}
	\label{fig:UglyZBP_18a}
\end{figure*}

\begin{figure*} % "Bad" ugly
	\includegraphics[scale=0.41]{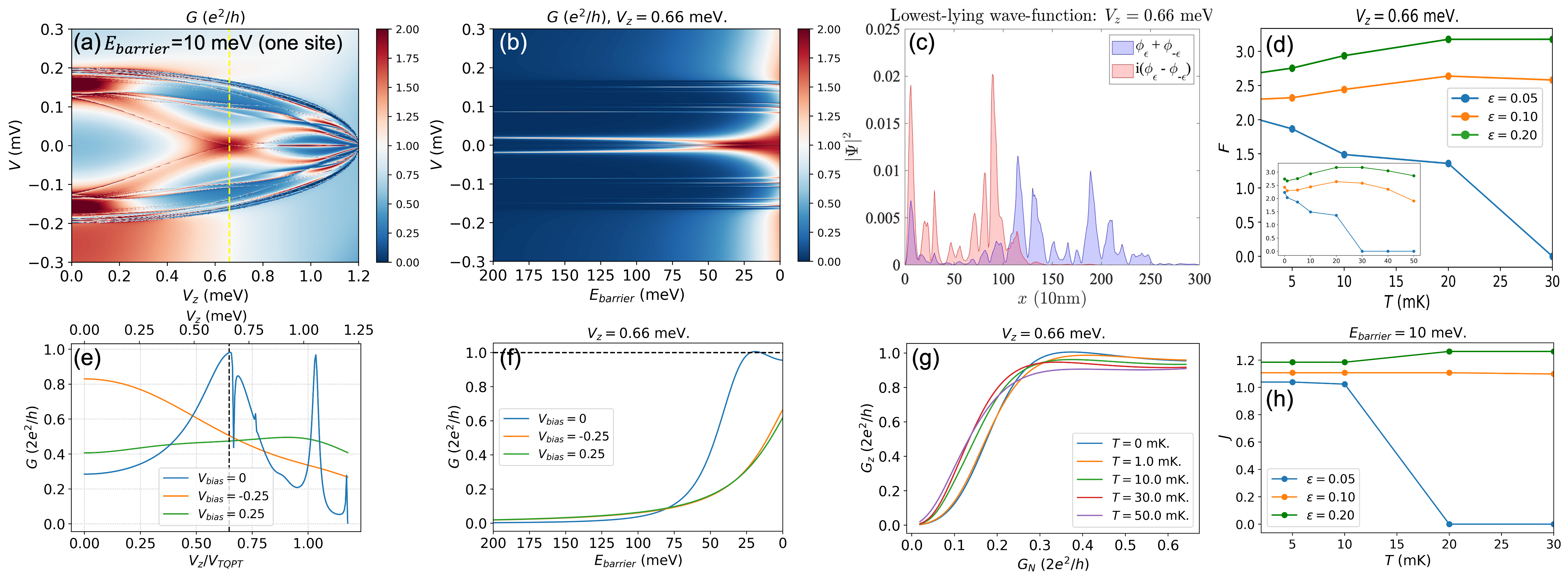}
	\caption{Numerical results for the ``ugly" ZBCP. Parameters: $\alpha=2.5$ meV, $\Delta_0=0.2$ meV, $V_c=1.2$ meV, $L=3.0$ $\mu$m, $\mu=1.0$ meV, $\lambda=0.2$ meV, and $\gamma=10^{-4}$ meV. The TQPT field is at $V_{\text{TQPT}}=\sqrt{\lambda^2+\mu^2}=1.020$ meV. The parameter for the on-site random potential is: $\sigma_\mu=2.0$ meV. (a) Conductance false-color plot as a function of bias voltage $V$ and Zeeman field $V_z$, with the fixed tunneling barrier height at $E_{\text{barrier}}=10$ meV. (b) Conductance false-color plot as a function of bias voltage $V$ and tunneling barrier height $E_{\text{barrier}}$, with the fixed Zeeman field at $V_z=0.66$ meV. (c) Lowest-lying wave-function probability density $|\Psi|^2$ as a function of nanowire position $x$, with fixed $V_z=0.66$ meV and $E_{\text{barrier}}=10$ meV. (d) Quality factor $F$ as a function of temperature $T$ for three different tolerance factors $\epsilon$, with the fixed $V_z=0.66$ meV. The inset figure gives an overall trend starting from $T=0$. At $T=10$ mK, $F=1.490$ for $\epsilon=0.05$. At $T=20$ mK, $F=2.638$ for $\epsilon=0.1$. (e) Conductance line cuts as a function of Zeeman field $V_z$ for three different bias voltages $V_{\text{bias}}$, with the fixed $E_{\text{barrier}}=10$ meV. The black dashed line marks $V_z=0.66$ meV. (f) Conductance line cuts as a function of tunneling barrier height $E_{\text{barrier}}$ for three different bias voltages $V_{\text{bias}}$, with the fixed $V_z=0.66$ meV. (g) Zero-bias conductance $G_z$ as a function of normal-metal conductance $G_N$ for five different temperatures $T$, with the fixed $V_z=0.66$ meV. (h) Quality factor $J$ as a function of temperature $T$ for three different tolerance factors $\epsilon$, with the fixed $E_{\text{barrier}}=10$ meV.}
	\label{fig:UglyZBP_18}
\end{figure*}
	
\begin{figure*} % "Good" ugly
	\includegraphics[scale=0.41]{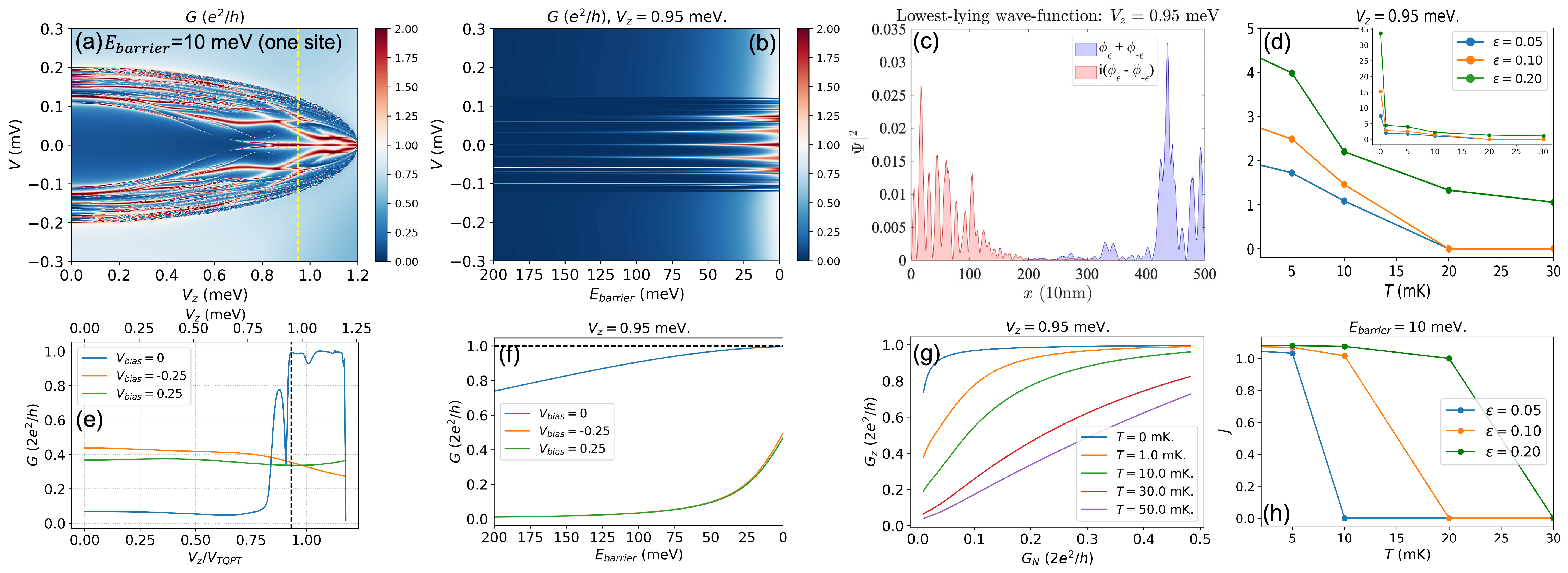}
	\caption{Numerical results for the ``ugly" ZBCP. Parameters: $\alpha=2.5$ meV, $\Delta_0=0.2$ meV, $V_c=1.2$ meV, $L=5.0$ $\mu$m, $\mu=1.0$ meV, $\lambda=0.2$ meV, and $\gamma=10^{-4}$ meV. The TQPT field is at $V_{\text{TQPT}}=\sqrt{\lambda^2+\mu^2}=1.020$ meV. The parameter for the on-site random potential is: $\sigma_\mu=1.0$ meV. (a) Conductance false-color plot as a function of bias voltage $V$ and Zeeman field $V_z$, with the fixed tunneling barrier height at $E_{\text{barrier}}=10$ meV. (b) Conductance false-color plot as a function of bias voltage $V$ and tunneling barrier height $E_{\text{barrier}}$, with the fixed Zeeman field at $V_z=0.95$ meV. (c) Lowest-lying wave-function probability density $|\Psi|^2$ as a function of nanowire position $x$, with fixed $V_z=0.95$ meV and $E_{\text{barrier}}=10$ meV. (d) Quality factor $F$ as a function of temperature $T$ for three different tolerance factors $\epsilon$, with the fixed $V_z=0.95$ meV. The inset figure gives an overall trend starting from $T=0$. At $T=10$ mK, $F=1.084$ for $\epsilon=0.05$. At $T=20$ mK, $F=0$ for $\epsilon=0.1$. (e) Conductance line cuts as a function of Zeeman field $V_z$ for three different bias voltages $V_{\text{bias}}$, with the fixed $E_{\text{barrier}}=10$ meV. The black dashed line marks $V_z=0.95$ meV. (f) Conductance line cuts as a function of tunneling barrier height $E_{\text{barrier}}$ for three different bias voltages $V_{\text{bias}}$, with the fixed $V_z=0.95$ meV. (g) Zero-bias conductance $G_z$ as a function of normal-metal conductance $G_N$ for five different temperatures $T$, with the fixed $V_z=0.95$ meV. (h) Quality factor $J$ as a function of temperature $T$ for three different tolerance factors $\epsilon$, with the fixed $E_{\text{barrier}}=10$ meV.}
	\label{fig:UglyZBP_15}
\end{figure*}
	
\begin{figure*} % "Good" ugly
	\includegraphics[scale=0.41]{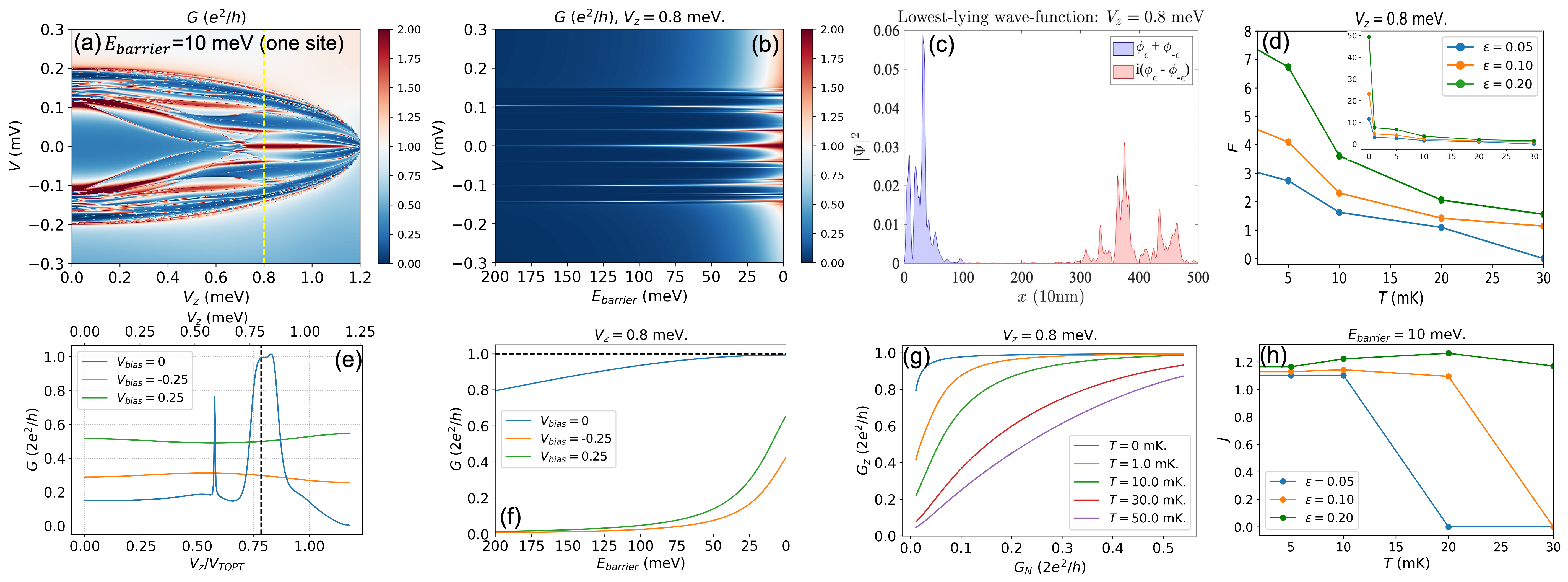}
	\caption{Numerical results for the ``ugly" ZBCP. Parameters: $\alpha=2.5$ meV, $\Delta_0=0.2$ meV, $V_c=1.2$ meV, $L=5.0$ $\mu$m, $\mu=1.0$ meV, $\lambda=0.2$ meV, and $\gamma=10^{-4}$ meV. The TQPT field is at $V_{\text{TQPT}}=\sqrt{\lambda^2+\mu^2}=1.020$ meV. The parameter for the on-site random potential is: $\sigma_\mu=2.0$ meV. (a) Conductance false-color plot as a function of bias voltage $V$ and Zeeman field $V_z$, with the fixed tunneling barrier height at $E_{\text{barrier}}=10$ meV. (b) Conductance false-color plot as a function of bias voltage $V$ and tunneling barrier height $E_{\text{barrier}}$, with the fixed Zeeman field at $V_z=0.8$ meV. (c) Lowest-lying wave-function probability density $|\Psi|^2$ as a function of nanowire position $x$, with fixed $V_z=0.8$ meV and $E_{\text{barrier}}=10$ meV. (d) Quality factor $F$ as a function of temperature $T$ for three different tolerance factors $\epsilon$, with the fixed $V_z=0.8$ meV. The inset figure gives an overall trend starting from $T=0$. At $T=10$ mK, $F=1.624$ for $\epsilon=0.05$. At $T=20$ mK, $F=1.419$ for $\epsilon=0.1$. (e) Conductance line cuts as a function of Zeeman field $V_z$ for three different bias voltages $V_{\text{bias}}$, with the fixed $E_{\text{barrier}}=10$ meV. The black dashed line marks $V_z=0.8$ meV. (f) Conductance line cuts as a function of tunneling barrier height $E_{\text{barrier}}$ for three different bias voltages $V_{\text{bias}}$, with the fixed $V_z=0.8$ meV. (g) Zero-bias conductance $G_z$ as a function of normal-metal conductance $G_N$ for five different temperatures $T$, with the fixed $V_z=0.8$ meV. (h) Quality factor $J$ as a function of temperature $T$ for three different tolerance factors $\epsilon$, with the fixed $E_{\text{barrier}}=10$ meV.}
	\label{fig:UglyZBP_19}
\end{figure*}
	
The last scenario considered in this paper is the case where a disorder potential of the type shown in Fig.~\ref{fig:Scheme}(d), as discussed in Sec.~\ref{sec:level1_2_2}, generates a ZBCP by chance as seen in Figs.~\ref{fig:UglyZBP_18a}-\ref{fig:UglyZBP_19}(a). Such ZBCPs would be present in only a small fraction of disorder realizations with the same parameters. These plots show results that span the set of possibilities, though their occurrence is somewhat rare. More generic results that do not show any quantization can be found in the Supplemental Material~\cite{SM}. All the cases of Figs.~\ref{fig:UglyZBP_18a}-\ref{fig:UglyZBP_19}(a) show conductance peaks that may be interpreted as gap closure or bound states~\cite{Pan2020Generic}. While a ZBCP arises in all these plots below the nominal TQPT at $V_z=V_{\text{TQPT}}$, Figs.~\ref{fig:UglyZBP_18a}-\ref{fig:UglyZBP_18}(a) show the gap closing feature merge into the ZBCP in a way similar to the cases of ``good'' and ``bad'' ZBCPs [i.e., Figs.~\ref{fig:GoodZBP_2}-\ref{fig:BadZBP_11}(a)], while Figs.~\ref{fig:UglyZBP_15}-\ref{fig:UglyZBP_19}(a) show a separation between the ZBCP and the gap closing peaks. Interestingly, the $V_{\text{bias}}=0$ line cuts shown in Figs.~\ref{fig:UglyZBP_18a}-\ref{fig:UglyZBP_19}(e) show that all the ZBCPs appearing in Figs.~\ref{fig:UglyZBP_18a}-\ref{fig:UglyZBP_19}(a) approach close to the quantized value in varying degrees. Figure~\ref{fig:UglyZBP_18a}(e) shows a somewhat broad plateau of ZBCP, which shows significant deviations from quantization that is quantified by the vanishing quality factor $J$ at $\epsilon=0.05$ plotted in Fig.~\ref{fig:UglyZBP_18a}(h). Figure~\ref{fig:UglyZBP_18}(e) shows quantized ZBCPs limited to small ranges, with a strong variation of the zero-bias conductance value between the quantized peaks that give rise to a very low quality factor $J\simeq 1.2$ even when the tolerance factor is $\epsilon=0.2$. Figure~\ref{fig:UglyZBP_15}(e) shows a nearly quantized plateau, which only approaches quantization in the vicinity of the TQPT field. Finally, Fig.~\ref{fig:UglyZBP_19}(e) shows a nearly quantized ZBCP, which is limited to a small range of Zeeman potentials that cannot be considered a plateau. Figures.~\ref{fig:UglyZBP_18a}-\ref{fig:UglyZBP_19}(b) show the variation of the height of the ZBCP for the case where the Zeeman field is chosen near the peak of the ZBCP (i.e., dashed line) in Figs.~\ref{fig:UglyZBP_18a}-\ref{fig:UglyZBP_19}(a) while the barrier height is varied. Figures.~\ref{fig:UglyZBP_18a}-\ref{fig:UglyZBP_18}(b) show that the ZBCP splits as the peak is reduced, revealing that the ZBCP was indeed non-topological, as can be confirmed from the Majorana decompositions shown in Figs.~\ref{fig:UglyZBP_18a}-\ref{fig:UglyZBP_18}(c). However, these ZBCP splittings in Figs.~\ref{fig:UglyZBP_18a}-\ref{fig:UglyZBP_18}(b) can merge and mimic zero-bias peaks when the finite-temperature effect kicks in, preventing ZBCPs to vary sharply as temperature changes, as in Figs.~\ref{fig:UglyZBP_18a}-\ref{fig:UglyZBP_18}(g), which are opposite to what we have seen in Figs.~\ref{fig:GoodZBP_2}-\ref{fig:BadZBP_11}(g). This is why we do not observe a sharp drop from the zero-temperature value of $F$ in Figs.~\ref{fig:UglyZBP_18a}-\ref{fig:UglyZBP_18}(d) to a finite-temperature one, and instead observe almost flat profiles of $F$ values as the temperature changes for $\epsilon=0.1$ and $\epsilon=0.2$. Due to their non-topological nature of the ZBCPs, the quality factors $F$ at $T=0$ are much smaller (below 3) in Figs.~\ref{fig:UglyZBP_18a}-\ref{fig:UglyZBP_18}(d) compared to other topological cases [Figs.~\ref{fig:GoodZBP_2},~\ref{fig:BadZBP_11},~\ref{fig:UglyZBP_15}, and~\ref{fig:UglyZBP_19}(d)]. With the apparent overlapping wave functions localized at one end in Fig.~\ref{fig:UglyZBP_18a}(c), the robustness of the ZBCP to changes in the tunnel barrier height cannot sustain at a practically measurable temperature of $20$ mK, which is quantified to be $F=0$ for $\epsilon=0.05$, indicating the complete lack of quantization for this parameter. On the other hand, the partially overlapping Majorana modes in Fig.~\ref{fig:UglyZBP_18}(c), which exhibit a bit more topological character compared to the previous case, reflect this part on the quality factor $F$---the value is $F=1.36$ at $T=20$ mK for $\epsilon=0.05$.
	
In contrast, Fig.~\ref{fig:UglyZBP_15}(b) shows the results of a disorder configuration where a ZBCP remains unsplit even under changes of the barrier height. This is consistent with the Majorana decomposition plotted in Fig.~\ref{fig:UglyZBP_15}(c), which shows a pair of separated Majoranas indicating a topological state. However, the rather strong delocalization of the Majorana wave functions implies that the tunneling conductance into these Majoranas will be rather weak. The result is that the ZBCP height plotted in Fig.~\ref{fig:UglyZBP_15}(g) turns out to be quantized only at temperatures below $10$ mK. This is a result of the fact that the nearly quantized ZBCP in Fig.~\ref{fig:UglyZBP_15} occurs only slightly below the nominal TQPT, i.e., $V_z\sim V_{\text{TQPT}}$. Thus, the result shown in Fig.~\ref{fig:UglyZBP_15} is better interpreted as the result of a reduction of the critical Zeeman field to reach the TQPT field as a result of the disorder potential. While disorder typically tends to suppress the topological phase and increase $V_{\text{TQPT}}$, it has been known to reduce the TQPT in rare fluctuations~\cite{Adagideli2014Effects}. Finally, the ZBCP height for the disorder configuration plotted in Fig.~\ref{fig:UglyZBP_19}(g) shows a plateau that appears almost as robust as the topological case shown in Fig.~\ref{fig:GoodZBP_2}. This is quantified by the value of the quality factor $F$ plotted in Fig.~\ref{fig:UglyZBP_19}(d) at $T=20$ mK, $\epsilon=0.05$ being closer to that of the ideal case shown in Fig.~\ref{fig:GoodZBP_2} at $T=20$ mK. The nearly topological behavior seen from ZBCP height can be cross-checked from the Majorana decomposition shown in Fig.~\ref{fig:UglyZBP_19}(c), which shows that the left Majorana wave function is spatially separated from the right wave function and should therefore be considered to be in the topological regime. One should note, however, that Fig.~\ref{fig:UglyZBP_19} shows a disorder configuration where the right Majorana would not be accessible to tunneling and will therefore fail the test of end-to-end conductance correlation between zero modes. This is, however, not a problem for several schemes for quantum computation~\cite{Vuik2019Reproducinga}.	

\section{Discussion}\label{sec:level1_4}
	
\subsection{Topological characterization based on the quality factor $F$}\label{sec:level1_4_1}
Let us now assess when the measured values of the quality factor $F$ can allow us to distinguish the case of topological (i.e., ``good'') MZMs  from the ``bad'' and ``ugly'' ZBCPs that can arise from end quantum dots and disorder. If we focus on the quality factor $F$ in the most ideal case,	which is set by choosing the smallest value $\epsilon=0.05$ and the lowest temperature $T\sim 0$ [i.e., insets in panel (d) of Figs.~\ref{fig:GoodZBP_2}-\ref{fig:UglyZBP_19}], we find that the ideal MZM is predicted to reach an $F$ in excess of 80, while this ideal value of the quality factors in the other cases (i.e., the ``bad'' and the ``ugly'' cases) remain below 20. This ideal value of the quality factor $F$, unfortunately, is not measurable in a finite-temperature experiment. The plots in panel (d) of Figs.~\ref{fig:GoodZBP_2}-\ref{fig:UglyZBP_19} show the quality factor over a more realistic range of temperatures. Looking back at Fig.~\ref{fig:GoodZBP_2}(d), we see that even measurements performed on an ideal MZM at a low temperature of 10 mK (i.e., $T=0.001$ meV) can only measure a quality factor of  $3.3$ for $\epsilon=0.05$, which is much below the $T\sim 0$ value of 84. The quality factor $F$ for the ``bad'' and the ``ugly'' cases are similarly reduced to be below 2 at $T\sim 10$ mK for $\epsilon=0.05$. The significant reduction of the quality factor at temperatures that are practical for transport measurements  potentially could make it difficult to distinguish between topological and non-topological MZMs. However, as already mentioned when discussing results for the ``bad'' and ``ugly'' cases, the good-bad-ugly paradigm is more of a classification of ZBCPs based on microscopic models rather than their ultimate topological characteristics. More specifically, an examination of the wave functions plotted in Figs.~\ref{fig:GoodZBP_2}-\ref{fig:UglyZBP_19}(c) reveals that many of the systems classified as ``bad'' and ``ugly'' based on models show spatially separated MZM wave functions that allow them to be classified as topological. While this occurs in more than one of the figures shown in this paper, these are rare for devices under the ``bad'' and ``ugly'' models, unless one specifically tunes the device to obtain a magnetic-field-stable ZBCP. One can find more generic examples of ``bad'' and ``ugly'' models with lower $F$ values that do not demonstrate separated wave functions in the Supplemental Material~\cite{SM}. In fact, of the results presented in this manuscript, only Figs.~\ref{fig:BadZBP_10}, \ref{fig:UglyZBP_18a}, and \ref{fig:UglyZBP_18} present truly non-topological results. These results show a quality factor $F$ below 1.5 at 10 mK for $\epsilon=0.05$. While the ideal topological ``good'' ZBCP in Fig.~\ref{fig:GoodZBP_2} gives $F=3.25$ at 10 mK for $\epsilon=0.05$, which is way above 1.5, a threshold of $F$ to distinguish topological ZBCPs can be carefully determined by the statistical distribution of the $F$ values for the non-topological ``bad'' and ``ugly'' cases (see the extended data in Supplemental Material~\cite{SM}). 

\begin{figure} 
	\includegraphics[scale=0.6]{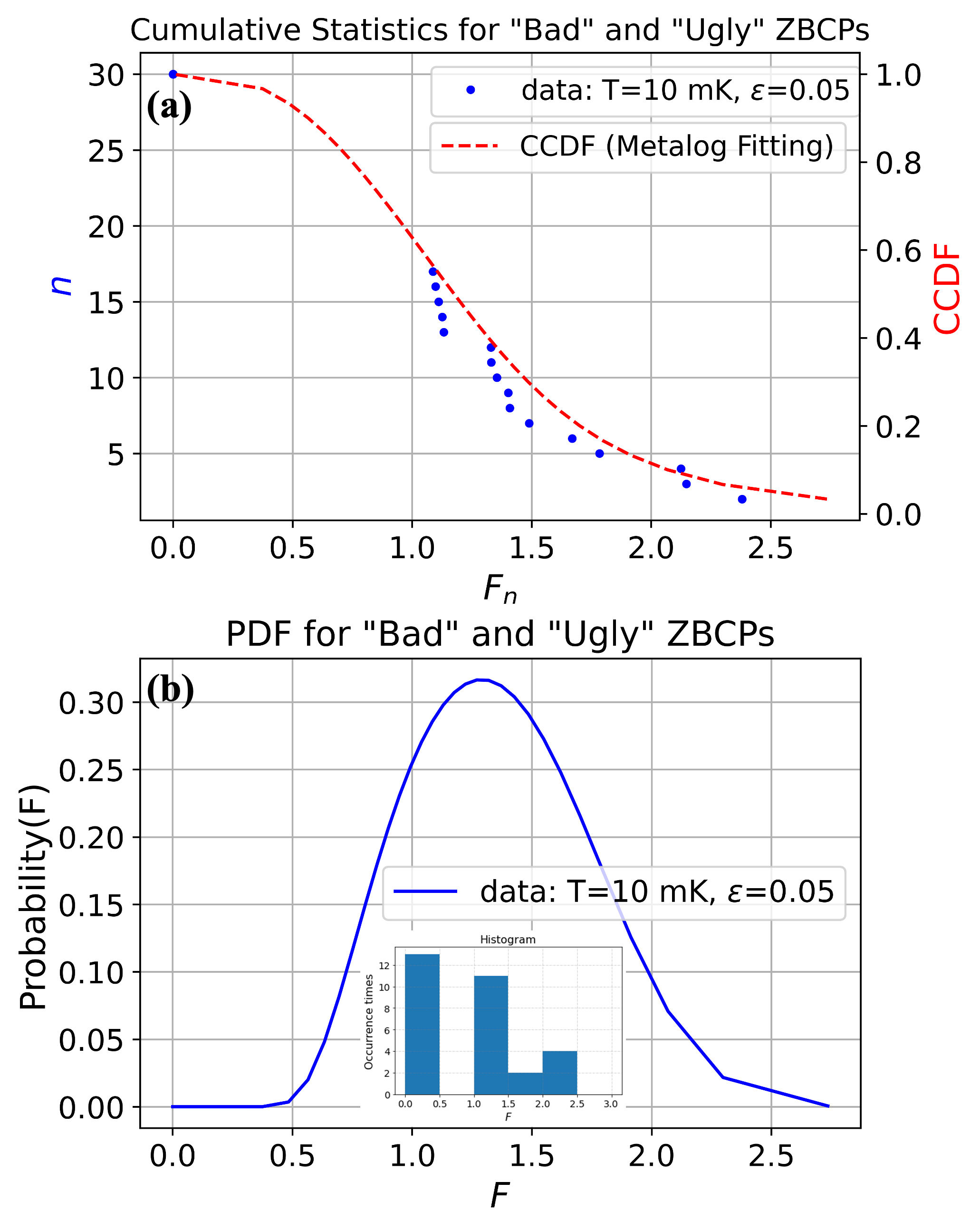}
	\caption{Statistical histogram for ``bad'' and ``ugly'' ZBCPs at $T=10$ mK and $\epsilon=0.05$. (a) Blue dots: Number of cases whose quality values $F_n$ ($n$ denotes each case) are larger than a given $F$ value on the $x$ axis. The data can be found in the Supplemental Material~\cite{SM}. Red dashed curve: Complementary cumulative distribution function (CCDF) is determined by fitting the blue dot data with a three-parameter metalog distribution. (b) Probability density function (PDF) is the (negative) derivative of the CCDF shown in (a). Inset: Regular histogram for the occurrence time with the bin width of 0.5 corresponding to the cumulative histogram in the main figure. The distribution from 0 to 1 is meaningless based on the definition of $F$ in Eq.~\eqref{F}.}
	\label{fig:histogram_T_10}
\end{figure}

\begin{figure} 
	\includegraphics[scale=0.6]{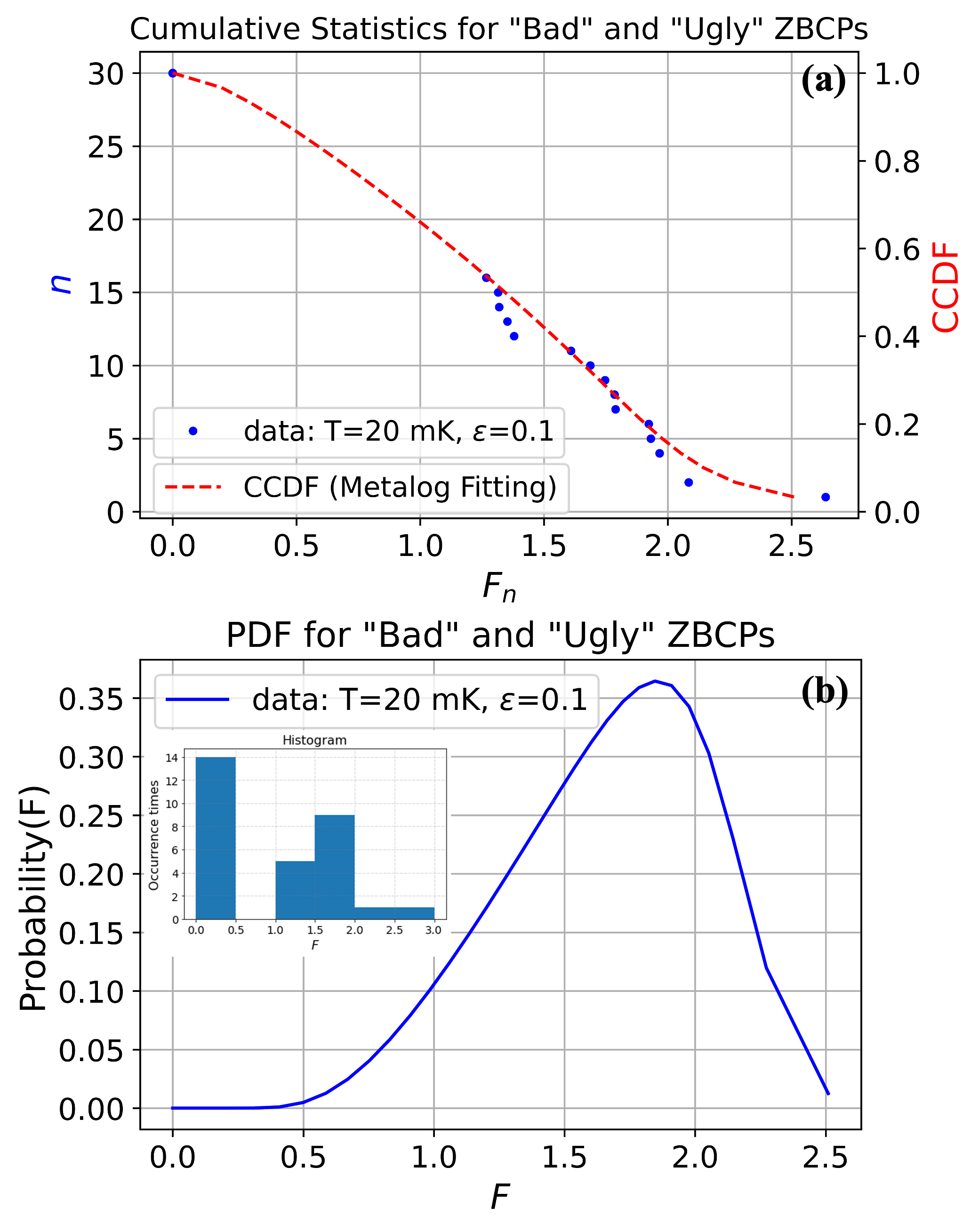}
	\caption{Statistical histogram for ``bad'' and ``ugly'' ZBCPs at $T=20$ mK and $\epsilon=0.1$.  (a) Blue dots: Number of cases whose quality values $F_n$ ($n$ denotes each case) are larger than a given $F$ value on the $x$ axis. The data can be found in the Supplemental Material~\cite{SM}. Red dashed curve: Complementary cumulative distribution function (CCDF) is determined by fitting the blue dot data with a three-parameter metalog distribution. (b) Probability density function (PDF) is the (negative) derivative of the CCDF shown in (a). Inset: Regular histogram for the occurrence time with the bin width of 0.5 corresponding to the cumulative histogram in the main figure. The distribution from 0 to 1 is meaningless based on the definition of $F$ in Eq.~\eqref{F}.}
	\label{fig:histogram_T_20}
\end{figure}

The results of $F$ values for the various models studied here can be summarized by the statistical distributions plotted in Figs.~\ref{fig:histogram_T_10} and ~\ref{fig:histogram_T_20}. The blue dots in Figs.~\ref{fig:histogram_T_10}(a) and~\ref{fig:histogram_T_20}(a) show the complementary cumulative distribution function (CCDF) from the values of $F$. This plot is obtained with $n$ along the $y$ axis and $F_n$ along the $x$ axis where $F_n$ are the $F$ values from the simulation listed in descending order. The CCDF at a particular $F$ represents the probability that a non-topological ZBCP can have an quality factor in excess of $F$. The red dashed curve in Figs.~\ref{fig:histogram_T_10}(a) and~\ref{fig:histogram_T_20}(a) are determined by fitting the CCDF using a three-parameter metalog distribution~\cite{Keelin2016Metalog} to the points with $F>1$. The corresponding probability density function (PDF), which provides a sense of the probability of obtaining a particular $F$ value and can be obtained from the fit to the CCDF is shown in Figs.~\ref{fig:histogram_T_10}(b) and~\ref{fig:histogram_T_20}(b). The insets of this figures show the simple histogram obtained from the $F$ data directly. However, it should be noted that the histogram from such a finite data set is strongly dependent on the bin sizes used to determine the histogram. The PDF obtained from the fits may be thought of as interpolations of the histogram.

Note that the data points in Figs.~\ref{fig:histogram_T_10}(a) and~\ref{fig:histogram_T_20}(a) either satisfy the constraint $F>1$ [follows from Eq.~\eqref{F}] for all cases where the ZBCP height reaches within a threshold $\epsilon$ of the quantized value or are set to $F=0$. The $F=0$ points represent the cases where the ZBCP remains below quantization as $F=0$. Though the data points in Figs.~\ref{fig:histogram_T_10}(a) and~\ref{fig:histogram_T_20}(a) contain a significant number of $F=0$ cases, these represent only a small fraction of the several hundred $F=0$ cases in our simulations. Specifically, to reduce the arbitrariness associated with sampling the high-dimensional parameter space in our model, we have restricted ourselves to parameters that can lead to a ZBCP whose height reaches within the threshold $\epsilon$ of the quantized value. For disordered samples (``ugly'' cases), this still leaves the possibility of disorder realizations having $F=0$ (i.e., not reaching quantization) even when other realizations for the same parameter have $F>1$. These are the $F=0$ cases contained in the plots of Figs.~\ref{fig:histogram_T_10}(a) and~\ref{fig:histogram_T_20}(a). The range of the dashed lines fitted to the CCDF for $F<1$ should be thought of as an interpolation to $F=0$. Since all $F<1$ represent ZBCPs below quantization and therefore are not candidates for a topological phase, the distribution of values of $0\leq F<1$ is irrelevant as long as the total probability in this range is preserved. 

Figure~\ref{fig:histogram_T_10}(a) shows that there is no non-topological case (``bad'' and ``ugly'') that exhibits a quality factor $F$ larger than 3, while the occurrence time for $F>2$ is already rare. With the extended data  of ``good'' ZBCPs, which all demonstrate $F>3$ at $T=10$ mK for $\epsilon=0.05$ (summarized on the first page of the pdf file of the Supplemental Material~\cite{SM}), the cumulative histogram for the trivial ZBCPs in Fig.~\ref{fig:histogram_T_10}(a) suggests a threshold of approximately $F=3$ would be sufficient to separate topological and non-topological ZBCPs. Based on the CCDF in Fig.~\ref{fig:histogram_T_10}(a), one can estimate the probability of a false positive, i.e., a non-topological ZBCP with a quality factor above $F$ would be less than $1\%$. While this estimate depends on how many $F=0$ cases we kept, one can be more conservative by restricting to only ZBCPs within the quantized range. In this case, since none of the 17 $F>1$ cases in Fig.~\ref{fig:histogram_T_10}(a) exceed the threshold, the probability of a false positive within the quantized range is less than $1\%$.

The demanding 10 mK temperature regime discussed in the last paragraph may be relaxed to 20 mK (which is close to the temperature in current experiments) by increasing the tolerance factor $\epsilon$ associated with $F$ to $\epsilon=0.1$. Similar to the analysis of the previous paragraph, using the statistical analysis in Fig.~\ref{fig:histogram_T_20}, we determine a threshold of $F=2.5$ for $\epsilon=0.1$ at $T=20$ mK. 
With this threshold, this false-positive case (i.e., the non-topological case whose quality factor exceeds the threshold) only accounts for $3.3\%$ of the data set of non-topological samples in Fig.~\ref{fig:histogram_T_20}(a), giving rise to a precision rate of $96.7\%$ to distinguish the non-topological ZBCPs from topological ZBCPs. This, while lower than the $T=10$ mK case, is still substantial, given that near-quantized ZBCPs are quite rare even in current experiments. To get a sense of this distinction, one can find that the ideal MZM shown in Fig.~\ref{fig:GoodZBP_2} is associated with an $F$ of $2.74$ for $\epsilon=0.1$ at $T=20$ mK, while all the presented non-topological cases [i.e., Figs.~\ref{fig:BadZBP_10}-\ref{fig:UglyZBP_19}] are below $1.8$, except for Fig.~\ref{fig:UglyZBP_18}. Considering that the quality factor in Eq.~\eqref{F} is defined to satisfy $F>1$, one should really consider $(F-1)>0$ as the true quality factor. With this redefinition, most of the data at $(F-1)< 0.8$ are almost a factor of 2 away from the threshold $F<2.5$.

Before we understand the outlier (i.e., case with $F>2.5$) in Fig.~\ref{fig:histogram_T_20}(a), detailed in Fig.~\ref{fig:UglyZBP_18}, let us consider the data that linger around $F=2.0$ (at $T=10$ mK for $\epsilon=0.05$) in Fig.~\ref{fig:histogram_T_10}(a), which demonstrate partially separated quasi-Majoranas that can be classified between being topological and non-topological. In short nanowires, which are most of the current experimental setups, MZMs appear as partially-separated quasi-Majoranas, which have been proposed to be able to implement braiding~\cite{Vuik2019Reproducinga}. Nonetheless, quasi-Majoranas are non-topological in the sense that the two Majorana modes would not localize ideally at both ends, which will fail the examination of the end-to-end conductance correlation~\cite{Lai2019Presence}. An example can be visualized in Fig.~\ref{fig:BadZBP_11}, which is labeled as a model of a ``bad'' ZBCP with a $F$ value of $2.15$ for $\epsilon=0.05$ at $T=10$ mK. Since Fig.~\ref{fig:BadZBP_11}(c) shows that the Majorana wave functions in this system are quite well-separated but inaccessible from the right end, this case would correspond to the quasi-Majorana scenario~\cite{Vuik2019Reproducinga}. Because of the separated MZMs, one can interpret this scenario as having a short segment of a topological superconductor at one end of the wire. 

The outlier in Fig.~\ref{fig:histogram_T_20}(a) corresponds to the ``ugly'' ZBCP results in Fig.~\ref{fig:UglyZBP_18}, which is associated with a value of $F=2.64$ for $\epsilon=0.1$ at $T=20$ mK. It is an example of the ``ugly'' ZBCP with a quality factor $F$ larger than the threshold that are mistakenly distinguished as topological, but does not possess topological characters based on its Majorana-decomposed wave functions (as we discussed in Sec.~\ref{sec:level1_3_3}). Similar to Fig.~\ref{fig:histogram_T_10}(a), one can find those cases~\cite{SM} associated with $F\simeq 2$ in Fig.~\ref{fig:histogram_T_20}(a) would exhibit partially-separated wave functions, known as quasi-Majoranas, while those cases associated with $F<1.5$ mostly show overlapping wave functions, which are clearly non-topological. So, the quality factor $F$ indeed demonstrates the quality of MZMs of different levels gradually. A high value of $F$ is seen as being associated with a MZM of good quality with sufficient confidence statistically.

This distinction between MZMs and quasi-Majoranas (or ABSs) is further blurred at $\epsilon=0.2$ where Figs.~\ref{fig:GoodZBP_2}-\ref{fig:BadZBP_11} show an $F$ value of approximately $5$ at $T=20$ mK. However, raising this tolerance $\epsilon$ seems to increase the separation between these MZMs/quasi-Majoranas and the ``ugly'' ZBCPs. In fact, the threshold for the quality factor $F$ depends on several factors, including the topological superconducting gap $E_g$ (which can be enhanced by increasing the spin-orbit coupling constant $\alpha$), nanowire length $L$, temperature $T$, and tolerance factor $\epsilon$. For a system with large $E_g$ and long $L$, the MZMs produced in the nanowires are expected to be quite robust against changes in the tunnel barrier height, so we can pick a high value of threshold, below which would be the non-topological cases. As the temperature $T$ is tuned lower, the MZM-induced ZBCPs will become closer to the ideal quantized plateau, giving a high value of $F$ (as we can see in Fig.~\ref{fig:GoodZBP_2}), while the ZBCPs induced from trivial subgap states can barely do the same (as we can see in Figs.~\ref{fig:BadZBP_10}-\ref{fig:UglyZBP_19}). The difference of $F$ values between topological and trivial ZBCPs can even become two orders of magnitude as the temperature approaches absolute zero. In this case, it is also easy to choose a higher value of threshold to separate the topological and non-topological ZBCPs. The tolerance factor $\epsilon$ can be viewed as the sensitivity to the quantization plateau. A smaller value of $\epsilon$ is only applicable when the ZBCP is closer to quantization, which can be realized by increasing the topological superconducting gap $E_g$, increasing the nanowire length $L$, and/or lowering the temperature $T$. Under these conditions, the gap of $F$ values between the topological and non-topological scenarios can be large, which makes us easy to choose the threshold of $F$. In contrast, when the experimental conditions do not facilitate the production of MZMs, the distinction of the $F$ values between topological case and trivial case can be comparatively minor, making it hard for choosing the threshold. This issue can also be seen from the false-positive case popping up in Fig.~\ref{fig:histogram_T_20} under the condition of $T=20$ mK with $\epsilon=0.1$ with a threshold of 2.5, while this issue is completely diminished in Fig.~\ref{fig:histogram_T_10} under the condition of $T=10$ mK with $\epsilon=0.05$ with a threshold of 3.0.

Among the ``bad'' and ``ugly'' results we present, the ones that demonstrate some kind of topological character (i.e., separated Majorana decomposed wave functions), such as Figs.~\ref{fig:BadZBP_11},~\ref{fig:UglyZBP_15}, and ~\ref{fig:UglyZBP_19}, in fact, do not pass the thresholds determined by the statistical histograms in Figs.~\ref{fig:histogram_T_10} and ~\ref{fig:histogram_T_20}. We can interpret these as the topological ZBCPs that cannot maintain their topological character in the finite-temperature environment when quantum dots or disorders interfere with the system. Since the threshold we set is to rigorously distinguish the ZBCP induced by MZMs, a measured value of $F$ higher than the threshold indicates a high probability ($> 95\%$) of getting a topological ZBCP. On the other hand, a value of $F$ below the threshold does not tell the scenario---it could be topological or non-topological. Therefore, our proposed quality factor $F$ can help filter out those ZBCPs with $F$ values larger than the threshold as potential MZM candidates to proceed with other Majorana examinations.

To apply this new metric $F$ to the experiments, we estimate the quality factor $F$ values through the recalibrated quantized conductance data from Zhang \textit{et al}.~\cite{Zhang2021Large}, which is our motivation to study the robustness of quantized ZBCPs. Under the experimental temperature 20 mK, the quantized plateau in Fig. 4(b) of Zhang \textit{et al.}'s paper~\cite{Zhang2021Large} gives rise to $F\simeq 1.16$ for $\epsilon=0.1$. Their extra data as shown in Fig. G2(b) also exhibits a quality factor value of $F\sim 1$ (for $\epsilon=0.1$ at 20 mK), which is way below $F=2.5$ as the suggested threshold in our simulation. These low values of $F$ demonstrating fragile quantization indicate the ZBCPs shown in Zhang \textit{et al.}'s paper~\cite{Zhang2018Quantized,Zhang2021Large} are very likely arising from non-topological subgap states.
	
\subsection{Relation to Majorana-based qubits}\label{sec:level1_4_2}
The estimates in the previous sub-section suggest that distinguishing MZMs from other non-topological ``bad'' and ``ugly'' ZBCPs might be rather experimentally challenging. This leads to questions about the motivation of using transport to make this distinction as opposed to time-domain techniques that work directly with nanowires in qubit configuration~\cite{vanZanten2020Photonassisted}. In this sub-section, we will argue that the quality factor $F$ provides a preliminary estimate of the decoherence rate for the MZM or the quasi-Majorana as a qubit. Alternatively, the best quality factor that is measured in a class of Majorana device provides a bound on how long of a coherence time one may expect for a Majorana qubit based on such devices. To understand this connection, let us assume the nanowire is long enough so that the overlap of Majoranas is smaller than the temperature being measured. In this regime, the estimated bit-flip rate of a topological qubit is determined by the topological superconducting gap $E_g$ and the temperature $T$ according to the relation $Te^{-E_g/2T}$~\cite{DasSarma2005Topologically}. The precision of quantization of the ZBCP associated with an MZM is limited by the same ratio $E_g/T$. More precisely, the width of ZBCP at $T=0$, $\Gamma$, in Eq.~\eqref{G0} is proportional to the normal-state conductance $G_N\propto \Gamma$~\cite{Setiawan2017Electron}. Accordingly, using the definition in Eq.~\eqref{F}, the quality factor $F$ can be approximated as
\begin{equation}\label{F_Gamma}
	F=\frac{G_{N,2}}{G_{N,1}}\sim\frac{\Gamma_{\text{max}}}{\Gamma_{\text{min}}}.
\end{equation}
The largest $\Gamma$ (controlled by tunneling) for which the ZBCP can be approximated as Lorentzian [i.e., Eq.~\eqref{G0}] is of the order of the topological gap, i.e., $\Gamma_{\text{max}}\sim E_g$.  For lower normal-state conductance, $G_N$, the conductance can only be quantized if $\Gamma_{\text{min}}\ge T$. Combining these arguments, we estimate that the quality factor for a ZBCP in a long Majorana wire with a topological gap $E_g $ at temperature $T$ to be $F\sim E_g/T$. From the previous sub-section, we concluded that within the class of models studied here, the quality factor $F$ needs to be greater than 2 to be relatively confident of a topological MZM. Assuming a qubit is build from an MZM of this quality, using the error rate estimate quoted earlier, we would conclude a rate of 56 MHz at 20 mK. This error rate is already higher than most non-topological qubits at the present point, which suggests that the threshold $F$ for meaningful topological qubit devices needs to be higher than that required to convincingly demonstrate a quantized ZBCP.
	
\subsection{Multi-channel effects and other Majorana systems}\label{sec:level1_4_3}
In this paper, for the sake of simplicity of presentation, we have restricted our results to single-channel semiconductor nanowires. However, as shown in the analysis of probing topological superconductivity through a quantum point contact~\cite{Wimmer2011Quantum}, the quantization of conductance into a multi-channel superconductor depends entirely on the number of channels in the tunnel contact, rather than the number of channels in the wire. The quantization of the topological conductance can thus be probed, as long as the tunnel contact can be gated by a tunnel barrier even if the number of channels in the topological superconduting nanowire is unknown, which is often the case. In fact, the tunnel barrier is typically tuned to be high such that the normal-state conductance is tuned to be below $2e^2/h$~\cite{Zhang2021Large}, suggesting that the quantum point contact is likely in the single-channel limit. This is because the transport studies of nanowires~\cite{Qu2016Quantized} have shown that the nanowire mobilities are large enough (i.e., ballistic transport) while the tunnel barrier is adiabatic potential (i.e., smooth Gaussian potential with the width larger than Landau Fermi wavelength) to show a few quantized steps as a function of gate voltage. Thus, even when the contacts on the two sides of the tunnel barrier are multi-channel, the transmission eigenvalues through the tunnel barrier are likely quite small in the higher channels. However, the small transmission in the higher channels may cause a small deviation from quantization of the topological conductance, which may be critical to explain observed conductances above $2e^2/h$~\cite{Nichele2017Scaling,Zhang2021Large}. This might be a limitation in our analysis of the relation between topological quantum computing and the quality factor from the last subsection, assuming that the conductance was in the tunneling limit, i.e., $\Gamma\ll \Delta$, which is equivalent to $G_N\ll G_0$. In this tunneling limit, it is reasonable to expect that the transmission of the tunnel contact dominantly involves a single channel. At the higher end of the tunnel conductance, i.e., $G_{N,2}$ in Eq.~\eqref{F_Gamma}, one can expect contributions from additional channels to cause deviations of the ideal MZM result from quantization. This can be avoided by ensuring that MZM conductance is measured through a clean segment of semiconductor nanowire, which can be controlled to be in the single-channel limit. This should be possible given the demonstration of clear conductance steps in transport through non-superconducting segments of such wires~\cite{vanWeperen2013Quantized,Qu2016Quantized}. An interesting future direction would be to understand to what extent the robustness of quantization can be used to establish multi-channel topological superconductors in other symmetry classes, which also appear to be associated with quantized conductance~\cite{BarmanRay2021Symmetrybreaking}.
	
Nearly quantized conductance has also been observed in STM measurements of vortices in iron superconductors~\cite{Zhu2020Nearly}. In this scenario, one can expect to be truly in the tunneling limit, where the tunneling process involves at most one channel near a few atom wide tip. A technical challenge involved in accurately measuring quantized conductance that is often overlooked is the small current involved in the measurement. To be specific, to measure a ZBCP of width $\Gamma$, which is near the lower end of the tunnel conductance used to estimate $F$, one needs to measure a current of the order of $I_{\text{ZBCP}}\sim \Gamma G_0 \sim k_B T G_0$. If one were to reduce the temperature to $20$ mK, this would amount to a current $I_{\text{ZBCP}}\sim 0.14$ nA. This is at the lower end of the currents measured in most experiments. The most general way to avoid this challenge as well as the requirement of going to temperatures as low as $20$ mK is to work with topological superconducting platforms with larger gap. Note that this does not simply refer to the topological superconducting gap, which is quite high for iron superconductors, but also the energy of sub-gap states such as vortex states in the case of iron SCs or low-energy ABSs at interfaces. 
	
\subsection{Characterizing MZMs based on $J$}\label{sec:level1_4_4}
Contrary to the quality factor $F$, which can be extremely large, the quality factor $J$ is limited. From the numerical results in Sec.~\ref{sec:level1_3}, $J$ values are all below $2.1$, which are comparatively lower when $F$ values can reach almost 50 in the zero-temperature limit. In general cases of ``good" ZBCPs, $J$ values could even be lower than those from ``bad" or ``ugly" ZBCPs when the Majorana splitting oscillations dominate. The best value of $J$ we can get for the ``good" ZBCP is $V_c/V_{\text{TQPT}}$ when the nanowire is longer than the SC coherence length and therefore the Majorana splitting is suppressed. The quality factor $J$ in this case turns out to be constrained by how soon the SC gap collapses and how late the system enters the topological regime. There is not much significant difference for the $J$ values between the ``good" ZBCPs and ``bad" or ``ugly" ZBCPs. Therefore, based on the models studied here, the magnetic field stability is difficult to use to characterize MZMs. However, a low value of the quality factor suggests a topological gap that may be too small for practical use.	
	
\section{Conclusion}\label{sec:level1_5}
In summary, we have studied the robustness of the ZBCP height relative to changes in the tunnel barrier height and magnetic field as a way to separate topological Majorana modes from trivial ``bad'' and ``ugly'' ZBCPs associated with subgap fermionic ABSs induced by inhomogeneous chemical potential and random disorder, respectively. This was motivated by the complete robustness to tunnel barrier height that is theoretically predicted for Majorana modes at zero temperature. In contrast to the experimental situation, theoretically, we have direct access to the Majorana wave functions and can determine in each case whether the system can be considered topological based on the spatial separation of the Majorana modes. While the magnetic field plateau, which we quantify as a quality factor $J$, is not a particularly strong indicator of the topological character of the system, the dimensionless quality factor $F$ introduced in Sec.~\ref{sec:level1_2_4} sharply quantifies the stability of the ZBCP to changes in normal-state conductance taking on values $\gg 80$ only in the topological phase. It should be noted that the quality factor $J$ is still a useful diagnostic to avoid a very small topological gap. By contrast, non-topological systems with strongly overlapping Majoranas show quantization over a very narrow range of normal-state tunneling conductance $G_N$, which typically leads to quality factors $F$ well below 10 at zero temperature. Therefore, the value of the low-temperature quality factor $F$ is a rather strong indicator of the topological character of the system. Unfortunately, for the realistic estimates of the topological gaps in currently existing semiconductor nanowire systems, our calculated quality factor $F$ at a somewhat realistic (albeit still challenging) temperature of $20$ mK turns out to be smaller even for the topological Majorana case discussed in Sec.~\ref{sec:level1_3_1}. Although this finding of ours for current nanowires is somewhat disappointing, it does not detract from the key role that the quality factor $F$ could play as a single diagnostic for the identification of emergent Majorana modes, particularly when improved materials fabrication enhances the topological gap. Even for small gaps, $F$ can distinguish between topological and trivial regimes, but perhaps not always decisively. The challenge should be alleviated by working with materials or systems with a larger topological gap. As discussed in Sec.~\ref{sec:level1_4_2}, this constraint for verifying the topological characteristics through transport is much softer than the constraint of realizing a topological qubit with an error rate comparable to existing nontopological qubit platforms. The additional challenge in this approach, as discussed in Sec.~\ref{sec:level1_4_3}, would be to design the tunnel barrier contact appropriately to ensure that the contact is actually in the single-channel limit. This might be a possible reason why robust quantization of Majorana conductance is still elusive. According to Wimmer \textit{et al.}~\cite{Wimmer2011Quantum}, a topological superconductor, when probed through a single-channel contact, is guaranteed to show quantized conductance over a range of tunneling conductance that is limited by temperature. This result does not depend on the number of channels of the nanowire~\cite{Wimmer2011Quantum}. Our results show that once these design, gap, and temperature constraints are achieved, finding a ZBCP with a quality factor $F$ in excess of 2.5 at 20 mK for $\epsilon=0.1$ (or in excess of 3.0 at 10 mK for $\epsilon=0.05$) should help establish a ZBCP as a topological Majorana mode. This criterion is consistent with $98.5\%$ of all the simulations (67 sets) presented in the Supplemental Material~\cite{SM} of which only seven are shown in the main text. Indeed, the significance of the quality factor $F$ should be interpreted in terms of the false positive probability at that value of $F$. But the false-positive rate for a non-topological case to have a high value of $F$ is very low---much less than $3.3\%$ as analyzed from our simulation results. In fact, no false positives were recorded for the 10 mK measurement at $\epsilon=0.05$. Thus, the small likelihood of false positives that this approach provides at the currently realistic parameters of $T=20$ mK vanishes if the temperature is lowered to $T=10$ mK or equivalently if the topological gap is doubled. Therefore, we believe that our proposed diagnostic $F$ should play an important role in all future topological superconducting platforms searching for non-local Majoana anyonic modes.

We note that in short wires, occasional quasi-Majorana modes do satisfy our $F>2$ criterion for Majorana stability because the wave-function overlap between the two MZMs forming the quasi-MZM happens to be small (thus leading to only one MZM of the pair coupling strongly with the lead). We believe, however, that such quasi-MZMs would behave, for all practical purposes, as topological MZMs, and should even enable successful braiding operations in short wires~\cite{Vuik2019Reproducinga}. After all, in finite length wires, there may not always necessarily be a difference between MZMs and quasi-MZMs because of finite wave function overlap.

We emphasize that although the importance of disorder in complicating the distinction between trivial and topological ZBPs was already known, the new physics in our work is to introduce a single dimensionless quantity $F$, which can distinguish between trivial and topological ZBCPs by analyzing the ZBCP stability in terms of experimentally measured quantities and not in terms of theoretical parameters. It is indeed true that in the end we find that the difference in $F$ between topological and trivial is rather small, but this unfortunately is the current reality in the existing nanowire systems. The currently available topological phase is extremely fragile because of disorder, and therefore any difference in $F$ between topological and trivial is not particularly large (i.e., about a factor of 2) for parameter estimates in current systems. We do not see any way out of this conundrum, but the positive aspect of our new physics is that the difference is measurable albeit small. With improving sample quality with less disorder or lower temperature, our quality factor $F$ will become a stronger indicator, distinguishing topological from trivial.

Acknowledgement: This work is supported by Laboratory for Physical Sciences and Microsoft. The authors acknowledge the support of the University of Maryland High Performance Computing Center for the use of the Deep Thought II cluster for carrying out the numerical work. 
	
\bibliography{Quality_ref.bib}
	
\clearpage
%\onecolumngrid
%\vspace{1cm}
\begin{center}
	{\bf\large Appendix}
\end{center}
%\vspace{0.5cm}
\setcounter{section}{0}
\setcounter{secnumdepth}{3}
\setcounter{equation}{0}
\setcounter{figure}{0}
\renewcommand{\thefigure}{A\arabic{figure}}
\newcommand\Scite[1]{[S\citealp{#1}]}
\makeatletter \renewcommand\@biblabel[1]{[S#1]} \makeatother
	
\section{Self-energy}\label{sec:levelA_1}
The self-energy $\Sigma(\omega,V_z)$ in Eq.~\eqref{H_NW} can be expressed as
\begin{equation}\label{selfE}
	\Sigma(\omega,V_z)=-\lambda\frac{\omega\tau_0+\Delta(V_z)\tau_x}{\sqrt{\Delta^2(V_z)-\omega^2}}\sigma_0,
\end{equation}
where $\lambda$ is the self-energy coupling strength with the parent SC. The self-energy is the effective renormalized energy introduced into the system when proximitized by the SC in the intermediate regime~\cite{Stanescu2010Proximity,Sau2010Robustness}. The Zeeman-field-varying SC gap is
\begin{equation}\label{SCgap}
	\Delta(V_z)=\Delta_0\sqrt{1-(V_z/V_c)^2}\cdot\theta(V_c-V_z),
\end{equation}
which hosts a bulk parent SC gap $\Delta_0$ without Zeeman field and vanishes above the SC collapsing field $V_c$. The Heaviside-step function $\theta(V_c-V_z)$ indicates that the proximitized SC effect no longer exists in the nanowire when $V_z>V_c$, which is not the interest of this paper. We will only numerically show the calculated conductance, energy spectrum, and wave functions below $V_c$.
	
The Hamiltonian in Eq.~\eqref{H_NW} becomes energy dependent when the self-energy $\Sigma(\omega,V_z)$ is included. Thus, to get the energy spectrum, instead of diagonalizing $H_{\text{NW}}(\omega)$ directly, we need to find the peaks of the density of states (DOS) located at energies $\omega_0$ from the Green's function, i.e.,
\begin{equation}\label{DOS}
	\rho_{\text{tot}}(\omega)=-\frac{1}{\pi}\text{Im}\left\{\text{Tr}\left[\tilde{G}(\omega)\right]\right\},
\end{equation}
and the Green's function is
\begin{equation}\label{GreemFn}
	\tilde{G}(\omega)=\frac{1}{\omega-H_{\text{NW}}(\omega)},
\end{equation}
where $\omega=\omega_0+i\eta$ and $\eta$ is an infinitesimal real number for DOS broadening. The trace function $\text{Tr}(\dots)$ in Eq.~\eqref{DOS} is over the spatial space and sub-space (i.e., particle-hole and spins) of the Green's function matrix $\tilde{G}(\omega)$. Note that all the numerical results in this paper include the self-energy because this is close to the real experimental situations. 
	
\section{Discretizing Hamiltonian}\label{sec:levelA_2}
To implement the numerical calculation, we have to discretize the continuum Hamiltonian as in Eq.~\eqref{H_NW} into a lattice chain of a tight-binding model~\cite{DasSarma2016How} with the lattice constant $a=10$ nm. Then the effective tight-binding tunneling strength $t=\hbar^2/(2m^* a^2)\approx 25$ meV. The effective spin-orbit coupling strength is $\alpha=\alpha_R/(2a)$. The length of the nanowire is given by $L=Na$, where $N$ is the total number of the atoms constituting the nanowire.
	
\section{Hamiltonian of NS junction}\label{sec:levelA_3}
The Hamiltonian of the normal lead is
\begin{equation}\label{H_lead}
	H_{\text{lead}}(\omega)=\left(-\frac{\hbar^2}{2m^*}\partial_x^2-i\alpha_R\partial_x\sigma_y-\mu+E_{\text{lead}}\right)\tau_z+V_z\sigma_x
\end{equation}
with the on-site energy $E_{\text{lead}}\approx -25$ meV in the lead controlled by the gate voltage. A tunnel barrier at the interface of the NS junction is modelled as a boxlike potential
\begin{equation}\label{TunnelBarrier}
	V_{\text{barrier}}(x)=E_{\text{barrier}}\Pi_{l_{\text{barrier}}}(x)
\end{equation}
in the barrier Hamiltonian
\begin{equation}\label{H_barrier}
	\begin{aligned}
		H_{\text{barrier}}(\omega)=&\left(-\frac{\hbar^2}{2m^*}\partial_x^2-i\alpha_R\partial_x\sigma_y-\mu+V_{\text{barrier}}(x)\right)\tau_z\\
		&+V_z\sigma_x-i\Gamma
	\end{aligned}
\end{equation}
to describe the interfacial scattering effect. $E_{\text{barrier}}$ is the tunnel barrier height and $l_{\text{barrier}}$ is the tunnel barrier width. $l_{\text{barrier}}$ occupies only one site in all the numerical results of this paper.
	
\end{document}